# An Evolutionary Framework for Cultural Change: Selectionism versus Communal Exchange

Running head: Evolutionary Framework Culture

LIANE GABORA
University of British Columbia

Address for Correspondence:
Liane Gabora
Department of Psychology, University of British Columbia
Okanagan Campus, Kelowna BC
Canada, V1V 1V7
E-mail: liane.gabora@ubc.ca
Phone: 604-822-2549 or 250-807-9849

**Abstract:** Dawkins' replicator-based conception of evolution has led to widespread mis-application of selectionism across the social sciences because it does not address the paradox that inspired the theory of natural selection in the first place: how do organisms *accumulate* change when traits acquired over their lifetime are *obliterated?* This *is* addressed by von Neumann's concept of a *self-replicating automaton* (SRA). A SRA consists of a self-assembly code that is used in two distinct ways: (1) actively deciphered during development to construct a self-similar replicant, and (2) passively copied to the replicant to ensure that *it* can reproduce. Information that is *acquired* over a lifetime is *not* transmitted to offspring, whereas information that is *inherited* during copying *is* transmitted. In cultural evolution there is no mechanism for discarding acquired change. Acquired change can accumulate orders of magnitude faster than, and quickly overwhelm, inherited change due to differential replication of variants in response to selection. This prohibits a *selectionist* but not an *evolutionary* framework for culture and the creative processes that fuel it. The importance non-Darwinian processes in biological evolution is increasingly recognized. Recent work on the origin of life suggests that early life evolved through a non-Darwinian process referred to as *communal exchange* that does not involve a self-assembly code, and that natural selection emerged from this more haphazard, ancestral evolutionary process. It is proposed that communal exchange provides an evolutionary framework for culture that enables specification of cognitive features necessary for a (real or artificial) society to evolve culture. This is supported by a computational model of cultural evolution and a conceptual network based program for documenting material cultural history, and it is consistent with high levels of human cooperation.

## 1. Introduction

Like organisms, elements of culture exhibit descent with modification; new ideas and artifacts build adaptively on previous ones. If it were possible to root the social and behavioral sciences in an evolutionary framework, they might achieve a unification comparable with Darwin's unification of the life sciences. Thus it is unsurprising that, dating back to Herbert Spencer's introduction of the notion of social Darwinism a few years after Darwin's *Origin of Species*, Darwinian thinking has been applied to a range of phenomena outside of biology, including creativity [14, 111], neural copying and pruning [11, 12, 13, 23, 26], law [55], cosmology [118], computer-mediated communication [74], and perhaps most extensively, cultural and economic change [7, 8, 15, 71, 86, 87, 107, 108]. Elements of culture build on one another cumulatively, as demonstrated even in laboratory settings [10]. Not only does culture accumulate over time, but it adapts, diversifies, becomes increasingly complex, and exhibits phenomena observed in biological evolution such as niches, drift, epistasis, and punctuated equilibrium [5, 22, 29, 32]. Processes of both biological and cultural evolution tend to gravitate toward a balance between differentiation (or divergence) and synthesis (or convergence) of different forms [98, 99]. Moreover, like biological evolution, culture is open-ended; there is no limit to the variety of new forms it can give rise to.

Clearly, those who participate in the evolution of culture are biological beings, but those who apply evolutionary principles to culture claim that culture constitutes a *second* Darwinian process, which although it piggybacks on the first, cannot be reduced to biology. It is acknowledged that some of what is considered cultural behavior *can* be accounted for by a purely biological explanation [4]. However, much as principles of physics do not go far toward an explanation of, say, the peacock's tail (though the physics of light and color play some role), biology does not go far toward an explanation of, say, the form and content of a haiku (though factors such as selective pressure for intelligence play some role).

To explain how and why such forms arise, accumulate, and adapt over time, it is necessary to develop a scientific framework for cultural evolution. Some cultural evolution research addresses specific cultural phenomena, such as cultural group selection [7] and cultural altruism [6, 30]. Such phenomena are generally tacked onto a selectionist framework, according to which culture evolves through a process algorithmically equivalent to natural selection. This paper summarizes why a selectionist theory of cultural evolution is inappropriate, and presents an alternative evolutionary framework for culture based on evidence that natural selection emerged from a more ancestral process of communal exchange. The paper aims to develop a theory of cultural evolution that can (1) explain why humans alone have evolved complex, cumulative, open-ended culture, (2) propose specific features of humans that enabled culture to evolve, and (3) predict whether new life forms, natural, artificial, or found elsewhere in our universe, should be able to evolve culture on the basis of these features. Selectionist accounts of cultural evolution have not yielded widely acceptable solutions to these challenges, nor is this accomplished by everyday observations of cultural change. Although it is undisputed that psychological and social phenomena are deeply *affected by* organic evolution, this paper focuses on applications of evolutionary theory to processes *other than* organic evolution.

*1.1. Terminology*
To make writings on these matters less awkward, the terms 'Darwinian' and 'selectionist' are used as a shorthand for 'by means of natural selection or a process that is algorithmically equivalent to it'. It would be wrong to interpret this as implying that Darwin never gave thought to evolution by means other than natural selection. He was, of course, immersed in the views of his day, and considered several possible explanations for adaptive change (*e.g.*, his ultimately unsuccessful theory of gemmules). Moreover, although Darwin did not use the term neutral evolution, he acknowledged that evolutionary change can involve fixation of variants that confer no selective advantage over previous adaptations; nevertheless, neutral evolution is commonly referred to as non-Darwinian [70, 75, 123]. Similarly, although Darwin was not committed to the idea that all life evolved from a single common ancestor, processes such as horizontal gene transfer (that is, genes transmitted between organisms in a manner other than through traditional reproduction) are commonly referred to as non-Darwinian [131]. Finally, the existence of non-Darwinian processes does not negate the feasibility of Darwinian ones, and *vice versa*. (For example, in biology, genes transferred horizontally can be fixed in a population by selection. In short, to facilitate the readability of this paper, 'Darwinism' is used in reference to, not the entire collection of musings Darwin had about evolution, but the theory he is famous for: natural selection.

## 2. The algorithmic structure of natural selection

The paradox faced by Darwin was: how do species *accumulate* change when traits acquired over their lifetimes are *obliterated?* His insight was to shift the focus from the organism to the population, and the distinction between inherited and acquired traits. Although acquired traits are discarded, inherited traits are not. When random mutations of inherited traits are beneficial for their bearers, their bearers have more offspring, and these traits proliferate at the expense of less beneficial ones. Over generations, this process of natural selection can lead to substantial change in the distribution of traits across the population.

Subsequently von Neumann showed that the minimal algorithmic structure capable of evolving through natural selection is a self-assembly code that gets used in two distinct ways [131]. First, it is actively deciphered (by what he referred to as the *constructor*) to construct a new automaton, a highly self-similar replicant of itself. Here, the code functions as rrrreeedreweedsdsebbhhj *interpreted information*. Second, it is used (by what he referred to as the *duplicator*) as a self-description that is passively copied to the replicant to ensure that *it* can reproduce. Here, the self-assembly instructions function as *un-interpreted information*. He referred to a structure that evolves through natural selection as a *self-replicating automaton* (Fig. 1).

[Insert Figure 1 here.]

Building on the concept of a self-replicating automaton, Holland [68] developed the genetic algorithm, a computer program that embodies the algorithmic structure of natural selection, and setting in motion the thriving field of natural computing (see [90] for a review). Holland proposed that structures, natural or artificial, that evolve through a selectionist process share three fundamental principles. The first principle is *sequestration of inherited information*; the self-assembly code is shielded from environmental influence. The second principle is a clear-cut *distinction between phenotype*, which is subject to acquired change, and *genotype*, which generally is not. The third principle is that natural selection incorporates not just a means by which *inherited* variation is *passed on*, but a means by which variation *acquired* over a lifetime is *discarded*. The rationale behind this last principle is simple. An organism can learn many things over the course of even a day, so acquired change can quickly exert a cumulative effect on an evolving lineage. But in order for a lineage to be affected by inherited change, an organism must mature and have offspring; inherited change works through a much slower process. It is only because, in biological evolution, changes due to the faster process (the effects of learning, aging, and so forth) are in general not retained from one generation to the next that changes due to the slower process (the effects of mutation and recombination) are not drowned out, and play an evolutionary role. In other words, since acquired change can accumulate exponentially faster than inherited change, if acquired change is *not* getting regularly discarded, it overwhelms Darwin's population-level mechanism of change; it 'swamps the phylogenetic signal'. Therefore, natural selection only works as an explanation if acquired change is negligible.

Darwin of course considered the problem of evolution from many angles, and considered many possibilities, including Lamarckian ones, but these were not part of his theory of natural selection, and are incompatible with it. In particular, he entertained the notion of pangenesis in a drastic attempt to explain such oddities as limb regeneration, hybridization, and atavism, which he was hard-pressed to make sense of without knowledge of modern genetics. However, it was clear to him that to the extent that pangenesis involves the transmission of acquired characteristics, adaptive change cannot be explained by natural selection, *i.e.,* by what is now referred to as a Darwinian mechanism. That is, he was aware that change *across generations* in the frequencies of *heritable variations* in a population due to selection would readily be drowned out by a mechanism for the transmission of *acquired traits* operating *within a single generation*.)

## 3. Efforts to provide a selectionist framework for cultural evolution

Having examined the algorithmic structure of natural selection, we can ascertain whether cultural evolution has this structure. Culture refers to the values, artifacts, customs, and so forth, of a social group. Elements of culture are transmitted vertically from one generation to the next, and horizontally amongst members of a generation. Thus, two key components of culture are: a means of generating novel behavior, and a means of spreading it through imitation and other forms of social learning.

A case can be made that thoughts and ideas constitute the cultural equivalent of a genotype, and that actions and artifacts constitute the cultural equivalent of a phenotype. One can argue that ideas are sequestered to the extent that others do not have access to an individual's thoughts. However, even speaking metaphorically, an idea or cultural artefact cannot be said to possess a self-assembly code that functions as both a passively copied self-description and actively interpreted self-assembly instructions [34]. There is no distinction between inherited and acquired information in culture. Artifacts such as musical scores and city plans contain *instructions*, but not *self-assembly instructions*; *i.e*., they do not generate more musical scores or city plans. Maturana and Varela [83] refer to an entity (*e.g*., a city plan) that generates another entity (*e.g*., a city) with an organization that is different from its own (*e.g*., a two-dimensional city plan generates a three dimensional city) as *allopoietic*. This is not the same thing as a self-replicating automaton that is composed of parts that through their interactions regenerate themselves and thereby reconstitute the whole, which Maturana and Varela refer to as *autopoietic*.

In sum, a selectionist explanation works only when non-random acquired change is either negligible, or assimilated into genetic self-assembly instructions as proposed by Waddington [134]. The strategic and potentially instantaneous modifications *acquired* by ideas as people think them through drowns out random changes occurring at the transition between generations, which in the case of culture means during transmission from one individual to another. Ideas routinely acquire change as they are put into one's own words and adapted to one's own tastes. In their externally expressed form, as actions or verbalizations, units of cultural information may appear not to have acquired change from one generation to the next; people re-deploy phrases, gestures, and problem-solving techniques that they have encountered before. However, they are often re-deployed in new situations, and moreover, in order for something to stick in memory we first relate it to what else we know, make it our own [100]. A Darwinian framework for culture is incompatible with this. It cannot begin to account for what happens when artists take familiar objects and themes and rework or juxtapose them, forcing us to adopt a richer appreciation of their meaning in different contexts.

Thus, a selectionist approach to culture must assume that humans are passive imitators and transmitters of pre-packaged units of culture, which evolve as separate lineages; it cannot accommodate non-random generation of variation and non-negligible transmission of acquired characteristics. Selectionist models of culture obscure this problem by emphasizing not the creative processes that generate cultural novelty but accidental errors or processes that bias the frequency of existing variants. Such processes include *conformity bias,* which occurs when people preferentially adopt widespread cultural variants over rare ones, *prestige bias,* which occurs when people preferentially imitate high status individuals, and *copying error* [7, 8, 61]. Darwinian approaches to cultural evolution such as memetics and Dual Inheritance Theory avoid incorporating any sort of creative or contemplative thought processes; ideas that exist in brains as a result of thinking something through for oneself have no place in their models: "[culture consists of] information stored in brains—information that got into those brains by various mechanisms of social learning…" [60, p. 120].

By limiting the psychological factors incorporated into Darwinian models to social learning, cultural Darwinists ignore those factors that exert the greatest impact on culture. These social learning factors tell us little about how cultural novelty arises. Indeed, creative individuals, those that make the most revolutionary contributions to culture, tend to be the least socially tethered of all, with strong leanings toward isolation, nonconformity, rebelliousness, and unconventionality [124]. Humans exhibit to a surprising degree the tendency to go our own way and do our own thing, and our artifacts reflect this. It was shown that creative works can be identified significantly above chance after they have been translated from one domain to another (*e.g.*, when artists are instructed to paint what a piece of music would look like if it were a painting, people identify which piece of music the painting was inspired by) [102]. Thus elements of human culture can be culturally rooted in seemingly unrelated sources. Moreover, our noteworthy achievements—the Mona Lisa, the steam engine, quantum mechanics—are widely known about and appreciated, but few attempt to imitate them, and it is not those who merely imitate them that make cultural history but those who deviate from what has come before[1].

It is not that transmission biases play no role at all in culture; what is at issue is the cultural Darwinist's focus on transmission bias. Transmission biases, while important, are less important to culture than the strategic, creative processes that generate and modify cultural content in the first place, and these creative processes cannot be accommodated by a Darwinian approach to culture because they entail retention of acquired characteristics. Since acquired change drowns out the slower inter-generational mechanism of change identified by Darwin (as explained earlier), natural selection is only of explanatory value when there is negligible transmission of acquired characteristics.

In contrast to social Darwinists, archaeologists and anthropologists who take an interpretive or social constructionist approach emphasize how cultural meanings emerge through interactions between individuals and their environments, emphasizing the impact of roles and perspectives, and applying agency theory to understand the relationships amongst individuals, societies, and institutions [66]. These approaches are better equipped to incorporate the strategic, creative processes underlying the generation of cultural novelty, though little attempt is made to develop an integrated scientific theory.

One reason a scientific theory of culture is difficult is that the paradox that *necessitated* the theory of natural selection does not apply to culture. Until the theory of natural selection was proposed, biologists could not explain how change accumulates given that, for example, if an animal is tattooed or loses its tail, its offspring are *not* born with tattoos or without tails. However, an analogous paradox does not exist with respect to culture; for example, once someone had the idea of putting a handle on a cup, cups could thereafter have handles. In biological systems, inheritance of novelty requires modification of the passively copied and actively transcribed genetic self-assembly code; in culture, there is no equivalent structure.

One might ask if analogous arguments hold for non-Darwinian aspects of biological evolution, and indeed they do. Because of phenomena such as mutualism, lineage reticulation (due to horizontal gene transfer and allopolyploidy—the combining the genomes of different parental species), certain traits evolve with astonishing speed, thereby diminishing the continuity and distinctiveness of species [53, 143]. Indeed, the stability of genetic information is so compromised that sequencing three *Escherichia coli* genomes revealed that fewer than 40% of the genes were common to all three [138]. As a result, the boundaries between many prokaryote species are fuzzy, and exhibit reticulate patterns of evolution, thus calling into question the appropriateness of the notion of the "tree of life" [64, 70]. The limitations of Darwinism as an explanation of the forms and dynamics of living things is increasingly recognized as the role of epigenetic processes becomes increasingly appreciated. Nevertheless, because such phenomena are much less present in biological evolution than cultural evolution, natural selection provides a reasonable approximation.

### *3.1. Replicators*

In *The Selfish Gene,* Dawkins sketched a simplified view of the gene-centric view of evolution pioneered by Williams [141, 142], Hamilton [56-58], and Maynard-Smith [84].

---

[1] As one choreographer (whose name I unfortunately forget) put it, if you're not doing what your predecessors did, you're doing what your predecessors did.

He claimed that biology is best understood by looking at the level of not the organism, nor the group, but the gene. According to Dawkins, a gene is *a replicator,* which he defines as "any entity in the universe which interacts with its world, including other replicators, in such a way that copies of itself are made" [19, pp. 17]. It is said to have the following properties:

- *Longevity*—it survives long enough to replicate, or make copies of itself.
- *Fecundity*—at least one version of it *can* replicate.
- *Fidelity*—even after several generations of replication, it is still almost identical to the original.

Replicators cause the entities that bear them to act in ways that make them proliferate. For example, genes cause individuals to act altruistically toward others such as close relatives who share those genes. Dawkins suggested that replicators can be found not just in biology but also in culture, and he termed these cultural replicators 'memes'. He claimed that much as genes propagate themselves in the gene pool by spreading from body to body via sperm or eggs, memes propagate themselves in the meme pool by spreading from brain to brain.

Clearly the principles of natural selection identified by Holland are not part of the replicator concept (Table 1). Unlike the notion of a self-replicating automaton, the replicator concept cannot explain why acquired traits are not transmitted across generations but inherited traits are. As such, it does not address the paradox of how species become increasingly adapted to their environments despite that modifications acquired by organisms during their lifetimes are not transmitted to offspring.

[Insert Table 1 here.]

Biologists expressed the concern that the replicator notion threw the baby out with the bathwater. It was occasionally dismissed as impoverished [74], or even "in complete conflict with the basics of Darwinian thought" [85]. However, it had a profound impact on the social sciences. Arguments against framing culture in Darwinian terms, and specifically in terms of memes, were mounted [28, 34, 39]. However, because of the popular success of *The Selfish Gene,* many social scientists were convinced that (a) replicators are the cornerstone of natural selection, and (b) natural selection is the only explanation for the accumulation of adaptive cultural novelty, it appeared self-evident that culture evolves through natural selection [3, 6]. After all, ideas stay roughly intact as they are transmitted from one individual to another; they appear to exhibit, to some degree, longevity, fecundity, and fidelity. Indeed the notion of a meme itself has been particularly proliferative; Derry [20] noted that in an internet search (http://www.google.com), 'meme*' beat 'gene*' by 45,500,000 to 28,100,000 pages, even after controlling for coincidences of the same word stem, foreign languages and peoples' names. Dual inheritance theorists went even further than memeticists, arguing that culture evolves through natural selection without so much as a replicator merely because it exhibits cumulative change[2] [60, 61].

---

[2] One might say, they not only threw the baby out with the bathwater but scoured away any dead skin cells clinging in the soap scum.

*3.2. Cultural representations are not particulate nor continuous but distributed*

It has been claimed that cultural representations are graded or continuous, and that because of this, cultural traits—unlike biological traits—can blend [1, 60].[3] There exist mathematical models that achieve blending by allowing cultural representations to take on continuous values. In one such model [60], each agent's belief is represented as a numerical value ($x$) between zero and one. They provide the following example: $x = 0$ represents the belief that the moon has emotions and that its color expresses its mood, while $x = 1$ represents the belief that the moon is simply a big rock without emotions, and its color is due to the laws of refraction. As an example of a belief that is intermediate between these extremes is the belief that the moon's color is 23% controlled by its emotions and 77% controlled by the laws of refraction, noting "such beliefs might seem odd to us because they violate intuitive expectations, which is why cognitive attractors might transform them" (p. 122). The beliefs $x = 0$ and $x = 1$ are referred to as *cognitive attractors* (following Sperber's [121] use of the term) because they are easiest to think. When they run the simulation they indeed find that the agents' beliefs shift over time from intermediate values to either 1 or 0. However, beliefs of the sort 'the moon's color is 23% controlled by its emotions and 77% controlled by the laws of refraction' are not just odd, they are psychologically unrealistic, and moreover the way people's representations generally change over time is not from complicated to simple but in the other direction, from simple to more complex, nuanced understandings of their world. Thus although it is claimed that the model improves our understanding of how culture evolves by paying close attention to the psychological processes involved [60, 61], it is unconvincing in this regard.

Others claim that cultural representations must be, at least in some sense particulate [3, 6, 30], as are genes, some models of cultural evolution assume particulate representations [*e.g.*, 7, 15]. The rationale here is that particulate models allow for retention of diversity; indeed it is because genes are particulate that genetic diversity is maintained across generations. If they were not particulate, variation would become diluted with time (much as mixing white and black paint gives gray paint; there is no way to go back to pure white or black paint). The argument is that a similar principle must be operating in cultural evolution.

However, although mental representations are not continuous, neither are they discrete and particulate [54, 89]. More precisely, at the neural level there is a sense in which representation involves discrete processes; a neuron either fires or it does not. But it is not because representations are continuous that they are subject to change, it is because they are *distributed* across cell assemblies of neurons that are sensitive to high level features or low level *microfeatures* [16]. For example, one might respond to edges of a particular orientation, or a sweet taste, or something that does not exactly match an established term [88]. Although each neuron responds maximally to a particular microfeature, it responds to a lesser extent to related microfeatures, an organization referred to as *coarse coding*. For example, neuron *A* may respond preferentially to a certain shade of blue, while its neighbour *B* responds preferentially to lines of a slightly different shade of blue, and so forth. However, although *A* responds *maximally* to one particular shade of blue, it responds somewhat to slightly different shades of blue. Thus the encoding of a mental representation is distributed across one or more cell assemblies containing many neurons, and likewise, each neuron participates in the encoding of many items [65]. The same neurons get used and re-used in different capacities, a phenomenon referred to as *neural re-entrance* [23]. Items stored in overlapping regions share features. Memory is also *content addressable;* there is a systematic relationship between the content of an input, and which particular neurons encode it[4]. As a result, items in memory can be evoked by stimuli that are similar or 'resonant' [59, 81].

[3] Actually, blending is possible with discrete traits. Indeed it is common in biology for polygenic traits (traits coded for by more than one gene), the classic example being Mendel's cross of red and white flowers to yield pink ones.

If there is overlap between regions in memory where two distributed representations are encoded, then they share one or more microfeatures. They may have been encoded under different circumstances, at different times, and the relationship between them never explicitly noticed. But because their distributions overlap, it is possible that some situation could cause one to evoke the other, thereby potentially modifying it. There are as many ways of generating associations as there are microfeatures by which they overlap; *i.e.*, there is room for atypical as well as typical associations. It is because the region of activated neurons is distributed, but not *too* widely distributed, that it is possible to generate a stream of coherent yet potentially creative thought. The more detail with which items have been encoded in memory, the greater their potential overlap, and the more routes for re-interpretting what is currently experienced in terms of what has been experienced before.

The bottom line is that it is not because representations are continuous that beliefs and ideas change over time, it is because they are distributed and content-addressable. Like particulate models, models based on distributed representations retain diversity, but like continuous models, they capture gradations. Moreover, they can begin to capture how ideas change when we think them through, combine them, and discuss them.

The distributed, content-addressable nature of memory is critically important for the creative processes by which representations acquire change [32, 42]. Associations between items are made, not by chance, but because the items share (high-level or low-level) features. Even if the two items were encoded at different times and the relationship between them never explicitly noted before, because they activated overlapping distributions of neurons, in a new context the relationship between them may suddenly become apparent. If memory were not distributed, there would be no overlap between items that share features, and thus no means of forging associations amongst them. If it were not content-addressable, associations would not be meaningful. Content addressability ensures the entire memory does not have to be searched or randomly sampled; one naturally retrieves items that are relevant to the current goal or experience, sometimes in a surprising but useful or appealing way. Actually, the notion of retrieving from memory is inaccurate; one does not retrieve so much as *reconstruct* [24]. This is why an item in memory is never re-experienced in exactly the form it was first experienced, but colored by what has been experienced in the meantime, and spontaneously re-assembled in a way that is relevant to the situation.

In short, much is known about how cultural representations are encoded in memory. However, this research is not encompassed by those who take a Darwinian framework. Due to the contextual, reconstructive nature of memory, representations non-randomly acquire characteristics between transmission events, which violates the conditions for such a framework to be applicable. By limiting the consideration of psychological processes to factors that bias 'who imitates who', a Darwinian explanation is not rendered invalid, because if one faithfully imitates ideas—if one never puts one's own slant on them—they do not acquire characteristics.

---

[4] Most computers are content addressable, but they work differently. Each different input is stored in a unique address. Retrieval is a matter of looking at the address in the address register and fetching the item at the specified location. Since there is no *overlap* of representations, there is no means of forging associations based on newly perceived similarities. The exception is computer architectures inspired by human associative memory. A connectionist memory is able to abstract a prototype, fill in missing features of a noisy or incomplete pattern, or create a new pattern on the fly that is more appropriate to the situation than anything it has ever been fed as input [65]. Like a human memory it does not retrieve an item so much as reconstruct it.

*3.3. A selectionist process requires random variation*

It has been argued that culture cannot be modeled as a process of natural selection because cultural variation is not generated randomly [60]. Pinker's [101] dismissal of the project is typical:

> "a complex meme does not arise from the retention of copying errors... The value added with each iteration comes from focusing brainpower on improving the product, not from retelling or recopying it hundreds of thousands of times in the hope that some of the malaprops or typos will be useful."

Others, however, argue that randomness is not necessary for the Darwinian framework to hold because "selective forces… require only variation not *random* variation" [60, p. 131].

The situation is in fact subtler than either of these positions. It *is* possible for a selectionist model to be applicable even if the underlying process is not random. However, although not genuinely random, the process must be approximated by a random distribution. When there are systematic deviations from a random distribution, natural selection is inapplicable as an explanatory framework because what is giving rise to change over time is the *nature* of those biases, *not* the mechanism Darwin identified: population-level change in the distribution of variants over generations of exposure to selective pressures. Biological variation is not genuinely random; for example, we can trace the source of some mutations to certain causal agents. However, the assumption of randomness generally holds well enough to serve as a useful approximation. Natural selection works by generating *lots* of possibilities through a process that can be approximated by a random distribution, such that at least one of them is bound by chance to be fitter than what came before.[5] Because the distributed, content-addressable architecture of human memory is specifically suited to obtaining *relevant* representations, in culture there is no need for large numbers of possibilities generated by chance processes. Culture works by intelligently generating *few* possibilities, such that they are more likely than chance to be adaptive.

In sum, for a selectionist theory to be applicable, there must be heritable variation with respect to traits that, if not random, can be approximated by a random distribution. To the extent that variation is biased away from random, what is giving rise to change over time is the nature of these biases, not selection. With respect to culture, not only is there no mechanism for inheritance (certainly not in the genetic sense, but not even in the abstract, algorithmic sense identified by von Neumann and Holland), but the processes that fuel cultural change are goal-driven, intuitive, strategic, and forward-thinking, *i.e.*, non-random.

---

[5] Actually, some biological situations, such as assortative mating, cannot be accurately approximated by a random distribution, and to the extent that this is the case natural selection gives a distorted model.

*3.4. Phylogenetic approaches to material cultural evolution*

It has become routine to use phylogenetic methods such as cladistics in an archaeological context [93-96, 106, 140]. In cladistic representations of archaeological data, the measured attributes of a 'taxon' of artifact are listed as a number string. The position in the string is loosely analogous to the concept of gene, and the number at that position is loosely analogous to the concept of allele. Thus if a taxon is represented by 132 then the first attribute is in state one, the second is in state three, and the third is in state two. For example, consider the representation of early projectile points from the Southeastern United States (Fig. 2) [93]. The data consist of metric and morphological measurements with respect to eight attributes, each of which can take from two to six possible states. Thus for example if fluting is absent in a particular artifact it has a 1 in position VII, and if fluting is present it has a 2. Seventeen 'taxa' are identified, and the pattern is such that one common ancestor (identified as KDR) gave rise to sequential branchings that culminated in 16 different taxa. This technique provides an intuitively meaningful (although potentially misleading) means of capturing structural change. The 'root taxon' at the far left is the most primitive, and early branch points represent changes that provided the structural constraints that shaped more recent changes. For example, much as evolution of the backbone paved the way for limbs, evolution of containers paved the way for spouts and handles.

[Insert Figure 2 here.]

Phylogenetic approaches have also been applied to culture in more complex ways. For example, relationships amongst different elements of culture have been analyzed by comparing their phylogenetic trees [67]. The procedure involves running a series of forward models, one in which the phenomena are assumed to evolve completely independently, another in which one kind of correlation is assumed, (*e.g.*, matriliny[6] with cattle), another in which a different correlation is assumed (*e.g.*, patriliny[7] with cattle). These are compared to the language phylogeny, which is assumed to be the most accurate available cultural history tree, to determine which gives the best match.

This method can indeed unearth relationships amongst different elements of culture. It was found, for example, that the spread of pastoralism in Sub-Saharan Africa is associated with a shift from matriliny to patriliny. The application of scientific tools and techniques to culture is a step forward, and these techniques work well for highly conserved assemblages of artifacts of one kind from one time period. However, despite their intuitiveness and scientific rigor, and some apparent successes, concerns have been raised about distortions generated by cultural applications of these methods [37, 47, 79, 125, 127, 130]. We now examine some of these concerns.

**Similarity Need Not Reflect Homology**. Phylogenetic methods assume that similarity reflects homology, *i.e.,* that two species are similar because they are related. Specifically, it assumes that, either (1) one is descended from the other, in which case shared traits were transmitted *vertically*, or (2) they are descended from a *common ancestor,* which is depicted as a branch point. For example, common ancestry can occur through *fission*, in which a population splits in two, which become increasingly differentiated.

---

[6] In matriliny, descent is traced through the mother's ancestry.

[7] In patriliny, descent is traced through the father's ancestry.

However, similarity need not reflect homology. Artifacts may arise independently yet be similar because they are alternative solutions within similar design constraints. This situation is referred to as *convergent evolution*, and it occurs in a biological context as well as in culture. However, because organisms must solve *many* problems (reproduction, locomotion, digestion, and so forth), the probability that a species is mis-categorized because of convergent evolution with respect to how it solves any *one* problem is low. Artifacts, on the other hand, are generally constructed with a single use in mind. (Though artifacts developed for use in one context are occasionally used to solve other problems, *e.g*., a screwdriver may be used to open a can of paint). Therefore, the probability of mis-categorization arising through the assumption that similarity reflects homology is problematic for artifacts. Tëmkin and Eldredge [125] showed that as a result of this and other related problems detailed below, a phylogenetic approach gives a distorted pattern of descent for the cornet and the Baltic psaltery, two musical instruments with well-documented ethnographic and archaeological records (Fig. 3).

[Insert Figure 3 here.]

**Predefined Attributes**. The data of Figure 2 are typical of those for which a phylogenetic approach is amenable, because the taxa are very similar to one another. That is, each taxon has one version or another of the considered attributes; there are no major modifications in this lineage. The units considered are those that are most amenable to analysis, which may differ substantially from those that were transmitted from teacher to apprentice. Units of actual cultural transmission must be communicable concepts, which are not necessarily conveniently measurable attributes. In other words, the method documents readily measurable change, not the actual cultural ancestry of the artifact. Relatedness between elements of culture often resides at the conceptual level, something not captured by phylogenetic methods, again due to their focus on measurable attributes. As a simple example, mortars and pestles are as related as two artifacts could be, despite little similarity at the attribute level.

**Blending**. Cultural relatedness frequently arises through not just vertical transmission but horizontal (inter-lineage) transmission, which can result in the blending of knowledge from different sources. Phylogenetic methods are ill-equipped to deal with inter-lineage transfer of information, in part because it is relatively rare in biological evolution compared to cultural evolution. They falsely classify similarity due to the horizontal exchange and blending of ideas as similarity originating from a common source [37, 125]. (Thus for example, it would assume that a cup with a handle with and a spoon with a handle were descended from a specific ancestral idea, though the similar feature could have arisen through horizontal transfer, *i.e.,* the handle on the spoon may have directly inspired the handle on the cup.) Extensive horizontal transmission or blending (whether biological or cultural) gives a bushy, reticulated appearance to a phylogenetic tree, which is misleading because it implies not just chronology but ancestry. Moreover, even if one is more interested in prehistoric culture than contemporary culture, one seeks not a bag of tricks for assessing relatedness each of which is applicable to certain data sets, but an explanatory framework that fits them all.

Another reason blending is problematic for selectionist-inspired methods is because they force one to parse the data according to predefined attributes or characters. They generally discard novel instances of blending because they do not fit cleanly into one predefined category or another. (Thus for example, in a collection of cups, if there was one with a handle it would not be included in the analysis.) One is *a priori* discouraged from incorporating data that does not fit into this parsing. In biology, such parsing arises naturally as a result of how traits are genetically encoded. The chosen attributes are characteristic of that species, and the rarity of inter-species mating ensures that they don't change drastically. However, in culture, nothing is *a priori* prohibited from 'mating with' anything else. Even the briefest examination of any modern household reveals that such blending is omnipresent. Blended artifacts exhibit the most reticulate (non-tree-like) evolutionary patterns, but they constitute what some would consider the most interesting, and certainly the most creative data we possess concerning our cultural evolutionary past.

**Lack of Objective Measure of Relatedness**. A more fundamental problem with phylogenetic approaches to culture is that they assume it is possible to accurately *measure* the relatedness of artifacts. Whether or not two organisms share a common ancestor is clear-cut; they either are or are not descendents of a particular individual. One can objectively measure what percentage of the genomes of two species overlap, and make conclusions about their degree of genetic relatedness. But in a cultural context, whether or not two artifacts "share a common ancestor" can be arbitrary, and moreover, what is measured is not necessarily what was culturally transmitted. For example, consider the following unusual artifact: a couch with a ledge-shaped seat and letter-shaped pillows that was inspired by the game of Scrabble (Fig. 4). Is this furniture descended from the game of Scrabble? Or was the Scrabble ledge which 'seats' a player's letters originally descended from the concept of a couch or chair? Did Scrabble somehow introduce a 'mutated taxon' into the 'furniture' lineage, or had the concept of an object to sit upon already contaminated the 'game' lineage when the inventor of Scrabble came up with the idea of perching letters on a ledge? What exactly was inherited?

[Insert Figure 4 here.]

In a world where thoughts of chairs can play a role in the invention of beanbag chairs, or scrabble ledges, or sketches of people seated on chairs, or even ballads about letters seated on scrabble ledges, how do you objectively determine what is or is not a variant or descendent of any particular idea? Is the article I read by you last year an ancestor of this one? Is the fitness of the Mona Lisa the number of prints made of it, or the number of people who looked at it for at least one minute, or who like it, or who imitated it, or who at least once tried to smile like the Mona Lisa? As Vetsigian *et al.* [131] put it, when ancestry arises through acquired change rather than inherited change, descent with variation is not genealogically traceable because change is not delimited to specific ancestors but affects the community as a whole. In the case of culture, a change to one concept or idea affects the 'ecology' of representations in an individuals' mind, which in turn affects how that individual responds to situations and contributes to the evolution of culture.

3.5. *Application of terminology from biology to culture*

The Darwinian view led not only to the mis-application of analytic techniques from biology to culture but to the mis-application of biological concepts. The fitness of a trait is generally measured in terms of the average reproductive success of individuals with that trait. However, the concept of fitness is occasionally applied to individuals (as opposed to traits) in a way that generates confusion when similarly applied in the cultural context. The *fitness* of an organism is the number of offspring it has in the next generation. In culture, there is no clear analog to biological fitness because ancestry is ambiguous.[8] A Darwinian approach is appropriate when it is possible to objectively measure the relatedness of individuals. Whether or not two organisms share a common ancestor is generally clear-cut; they either are or are not descendents of an ancestral organism. One can measure what percentage of the genomes overlap, and make conclusions about their genetic relatedness. For elements of culture, however, there is no objective way to go about this. The greater the extent to which information is exchanged horizontally, the more difficult it is to delineate the ancestors or *variants* of a particular individual.

It has been claimed that inheritance is a key Darwinian property of cultural as well as biological evolution [87], and some anthropologists use the term dual inheritance theory to refer to efforts to explain how human behavior is a product of both genetic and cultural evolution [8, 60, 61]. However, applying the term *inheritance* to culture is also problematic. That which is transmitted through culture is not encoded in self-assembly instructions, and therefore it is acquired, not inherited. Use of the term 'dual inheritance' to refer to both what is transmitted genetically and what is transmitted culturally is thus technically incorrect and misleading [43].

Applying the term *generation* to culture is similarly problematic. The term is applicable when individuals are irretrievably lost from a population and replaced by new ones. This is the case in a biological context, but not in a cultural context; an outdated (seemingly 'dead') idea can come back into use (seemingly 'come back to life') if styles change or circumstances become right. Since there is no clear distinction between a living entity and a dead one, there is no basis for determining what constitutes a generation. Thus the term 'generation' does not strictly apply to culture.

## 4. An evolutionary framework for creativity?

Not just cultural change but also the individual-level creative processes that fuel it have also been described as Darwinian. Ideas become more complex and adapted over time, even in the mind of a single creator before they are expressed to others [14, 30, 92, 126, 127]. Converging evidence suggests that creative ideation involves shifting between two forms of thought [27, 33, 69, 82, 135]: (1) *A divergent* or *associative* process that predominates during idea generation, and (2) a *convergent* or *analytic* process that predominates during the refinement, implementation, and testing of an idea. There is evidence that associative thought is facilitated by a positive mood, while analytic thought is facilitated by a negative mood [120]. The capacity to subconsciously shift between these modes depending on the situation, by varying the specificity of the activated cognitive receptive field has been referred to as *contextual focus*[9] because it requires the ability to focus or defocus attention in response to the context or situation one is in [33]. Defocused attention, by diffusely activating a broad region of memory, is conducive to divergent thought; it enables obscure (though potentially relevant) aspects of the situation to come into play. Focused attention is conducive to convergent thought; memory activation is constrained enough to hone in and perform logical mental operations on those aspects of a situation that are clearly relevant to the task at hand. Thus not only are the products of creative ideation indicative of an evolutionary process—*i.e.,* they exhibit cumulative complexity and adaptation over time—but we appear to possess cognitive processes that specifically operate at different stages of creative ideation. As in natural selection, one phase is conducive to generating variety, and another phase is conducive to pruning.

---

[8] I too have been guilty in the past of mis-applying the term fitness in a cultural context.

[9] In neural net terms, contextual focus amounts to the capacity to spontaneously and subconsciously vary the shape of the activation function, flat for divergent thought and spiky for analytical.

The theory that creativity in the mind of an individual is an evolutionary process was proposed by Donald Campbell [14], who claimed that we generate new ideas through 'blind' variation and selective retention (abbreviated BVSR). Today the most well known proponent of BVSR is Dean Keith Simonton [110-117]. As Simonton [110, p. 310] puts it:

> [H]ow does the individual arrive at new ideas in the first place? How do human beings create variations? One perfectly good Darwinian explanation would be that the variations themselves arise from a cognitive variation-selection process that occurs within the individual brain. Not surprisingly, one of the champions of this Darwinian theory of creativity was Donald Campbell (1960).

Like Campbell, Simonton views creativity as essentially a trial-and-error process in which the most promising 'blindly' generated ideational variants are selected for development into a finished product. One varies the current thought a multitude of different ways, select the fittest variant(s), and presumably this process is repeated until a satisfactory idea results. The variants are said to be 'blind' in the sense that the creator has no subjective certainty about whether they are a step in the direction of the final creative product.

Simonton distinguishes between *primary Darwinism*, which refers specifically to evolution of species through natural selection, and *secondary Darwinism*, which refers more generally to processes that possess the algorithmic structure of a selectionist process. He refers to creativity as an instance of secondary Darwinism, although he appears to believe that the two are not so distinct:

> I have argued here that secondary Darwinian theory, as proposed by Campbell (1960), furnishes the most complete basis for comprehending human creativity. … Campbell's secondary Darwinian model of creativity may be eventually subsumed under primary Darwinian theory ([110], p. 323).

Many problems with this theory have been laid out [18, 25, 35, 38, 122, 136, 137]. One problem with BVSR is that a selectionist process requires a population of simultaneously existing entities exposed to the same selection pressure (or adaptive landscape). But in a stream of creative thought, since each successive thought or idea (or version of an idea) alters how one conceives of the problem or task, no two are simultaneous, and thus no two are evaluated with respect to the same criteria or selection pressures. Thus they cannot be treated as members of the same generation as is necessary to select which are fittest.

Another indication that BVSR is on the wrong track is that, as was the case with cultural evolution, the problem for which natural selection was put forward as a solution does not exist with respect to creativity. That is, there is no sense in which characteristics or components of a creative idea cyclically accumulate and then get discarded at the interface between one generation and the next. For example, once a composer decides that a particular piece will be in the key of E, or that the bassoon will come in at bar 16, the piece generally does not revert back to the state of not having acquired these characteristics. One might ask if Darwin's solution is nevertheless applicable; might processes outside of biology evolve through a selectionist process even if natural selection was originally advanced as a solution to a paradox that is unique to biology? The problem, once again, is that since acquired change can accumulate orders of magnitude faster than inherited change, if it is *not* getting regularly discarded, it quickly overwhelms the population-level mechanism of change identified by Darwin. This is particularly problematic with respect to creative thought since novelty does not originate through random processes—which are prone to cancelling one another out—but through strategic processes (or implicit, intuitive processes, making use of the associative structure of memory)—which are not prone to cancelling one another out.

After referring to his theory for over a decade as the Darwinian theory of creativity, Simonton has recently backed away from a Darwinian conception of the creative process [117]. While still using the acronym BVSR, and he claims that his theory has been "radically reformulated" to show that "BVSR's explanatory value does not depend on any specious association with Darwin's theory of evolution". Nevertheless there is *some* sense in which creative ideas evolve over time.

There is increasing evidence in support of a non-Darwinian evolutionary theory of creativity, *honing theory*, according to which creative ideas evolve not by selecting amongst multiple blindly generated variants, as postulated by BVSR, but by actualizing the potential of an ill-defined idea, by observing how it changes state through interaction with sequentially considered contexts or perspectives [35, 49, 51, 103]. The ill-defined idea can be said to be in a 'state of potentiality' because it could actualize different days depending on the different perspectives or contextual cues taken into account as it takes shape. Honing theory is derived from the notion that individuals' internal models of the world, or *worldviews,* self-organize to achieve a more stable equilibrium state [31, 32, 44]. Creative outputs are viewed as the external byproducts of this internal, transformative, entropy-minimizing process; thus the theory is consistent with the therapeutic nature of the creativity [48].

## 5. Evolution through communal exchange

Growing evidence that the evolutionary dynamic that gave rise to translation was non-Darwinian [131] suggests that natural selection may be inapplicable to the early stages of *any* evolutionary process. Indeed in developing a theoretical framework for cultural evolution, it is instructive to look at how early life itself evolved. Research into the origin of life is stymied by the improbability of a spontaneously generated structure that replicates using a self-assembly code (such as the genetic code). This has led to widespread support for the hypothesis that the earliest protocells were not self-replicating automata but autocatalytic molecular networks that evolved (albeit haphazardly) through a non-Darwinian process involving horizontal (lateral) transfer of innovation protocols [36, 73, 91, 105, 131, 133, 143]. This process (illustrated schematically in Fig. 5) has been referred to as *communal exchange* [131].

[Insert Figure 5 here.]

In communal exchange, information is transmitted not by way of a self-assembly code from parent to offspring, but through interactions with the environment, including other protocells (Table 2). The protocell replicates through duplication of individual elements followed by budding or cell division; communally exchanged information may therefore be retained. Early life underwent a transition from a fundamentally cooperative process of horizontal evolution through communal exchange to a fundamentally competitive process of vertical evolution through natural selection by way of the genetic code. This transition is referred to as the *Darwinian threshold* [131]. It has been estimated that the period between when life first arose and the Darwinian threshold spanned several hundred million years (K. Vetsigian, Pers. Comm., November 24, 2008].

[Insert Table 2 here.]

Note that there was some degree of reproductive isolation in this primitive world, in the sense that some protocells could not interact with others, if only due to spatial segregation, and this could be expected to increase over time as 'protospecies' became increasingly differentiated. However, as is the case with respect to culture, a seemingly 'dead' protocell could 'come back to life' if the molecular environment changed and circumstances became right [36]. Again, because there is no hard and fast distinction between a living entity and a dead one, there is no basis for determining what constitutes a generation. Thus, as with cultural evolution, the term 'generation' does not apply to the earliest stage of biological life [36, 131].

## 6. Applying communal exchange to cultural evolution

If natural selection cannot explain how life arose, nor the earliest chapter of its evolutionary history, it is unsurprising that efforts to apply it to culture have yielded little in the way of explanatory or predictive power. If it took time for natural selection to emerge as the mechanism by which life evolves, it seems reasonable that culture too would evolve by way of this more primitive mechanism. Thus it is proposed that, like these earliest life forms, culture evolves through a non-Darwinian process of communal exchange. What evolves through culture is *worldviews,* the integrated webs of ideas, beliefs, and so forth, that constitute our internal models of the world, and they evolve, as did early life, not through competition and survival of the *fittest* but through transformation of *all* (Fig. 6). The assemblage of human worldviews changes over time not because some replicate at the expense of others, as in natural selection, but because of ongoing mutual interaction and modification. Rituals, artifacts, and other elements of culture reflect the states of the worldviews that generate them.

[Insert Figure 6 here.]

Application of a communal exchange or network-based approach to cultural evolution began with recognition of the fact that analyses of what sort of formal structure had to emerge to bring about organic evolution have been useful in thinking about what sort of formal structure a mind must have in order to participate in cultural evolution. Cross-fertilization between formal communal exchange models of these two evolutionary processes is still in its infancy, but there are indications that this is a fruitful avenue for future research.

An early effort along these lines [31, 32] was inspired by Kauffman's [73] application of graph theory to the emergence of autocatalytic sets of polymers. He recognized that theories of how life began must solve the "chicken and egg" problem of how a complex system composed of mutually dependent parts could come into existence. He showed that when polymers interact, their diversity increases, and so does the probability that some subset of the total reaches a critical point where there is a catalytic pathway to every member, a state Kauffman referred to as *autocatalytic closure*. He demonstrated that autocatalytic sets emerge for a wide range of hypothetical chemistries—*i.e.,* different collections of catalytic molecules. Some sets were *subcritical*—unable to incorporate new polymers—and others were *supracritical—*able to incorporate new polymers with each round of replication. Which of these two regimes a particular set fell into depended on the probability of any one polymer catalyzing the reaction by which a given other was formed, and the maximum length (number of monomers) of the original "food set" polymers.

The rationale for using an analogous approach to model the conditions for the emergence of cultural evolution is that once again we are faced with the problem of explaining how a complex system composed of mutually dependent parts could come into existence. In order to contribute in a meaningful and reliable way to cultural evolution, one must be able to improve on ideas by thinking them through, *i.e*., engage in abstract thought. However, abstract thought is the process that connects thoughts and ideas in the first place, *i.e*., that puts them within reach of one another. Kauffman's framework can be adapted to this new "chicken and egg problem", making plausible assumptions about associative memory [31, 32]. The analog of the set of polymers is the set of items encoded in associative memory. The analog of *M*, the maximum polymer length, is the maximum number of features of an item encoded in memory when one attends that item, and the analog of *P*, the probability of catalysis is the probability that one thought brings about associative recall of another. So long as exposure to highly similar items or events causes the formation of abstract concepts that connect these instances, an associative memory that meets certain criteria is expected, sooner or later, to reach a critical percolation threshold such that the number of ways of forging associations amongst items in memory increases exponentially faster than the number of items in memory. This is an entropy-minimizing process and the outcome is that the worldview achieves a more stable equilibrium state that (following Kauffman's use of the term 'autocatalytic closure') has been referred to as *conceptual closure* [31, 32, 44].

If the probability of associative recall is low, the network is subcritical. The resulting worldview will tend to be stable but may have difficulty incorporating new information. If the probability of associative recall is high, the network is supra-critical. The resulting worldview will incorporate new information readily but it but risks destabilization. In Kauffman's origin of life model, each polymer was composed of up to a maximum of *M* monomers, and assigned a low *a priori* random probability *P* of catalyzing each reaction. The lower the value of *P*, the greater *M* has to be, and *vice versa* in order for closure to be achieved. In conceptual closure, as was the case with autocatalytic closure, if *M* or *P* is too high the behavior of the network may be too chaotic to establish closure. The key point is that interactions amongst items in an associative memory increases their joint complexity, eventually transforming them into a conceptual network, which transforms as new inputs are incorporated. This enables the creative connecting and refining of concepts and ideas necessary for the individual to participate in the evolution of cultural novelty.

Note that prior to the Darwinian transition in the evolution of biological life, some molecular elements were more readily incorporated into protocells than others. In this sense one could say that competition is taking place, but it is competition at the level of the *components* of protocells. It is not competition at the level of the self-replicating automaton, the protocell itself; there is no competitive exclusion amongst instruction sets for building self-replicating structures such that some are selected at the expense of others. It is therefore not a sort of competition that is indicative of natural selection. Similarly, one might say that different styles or brands compete in the marketplace. However, this kind of competition in the external world does not imply competitive exclusion at the level of the underlying worldviews. Again, this is not competition at the level of self-replicating automata—there is no selection of some instruction sets for self-replicating structures over others—and thus it is not the sort of competitive exclusion that is indicative of a selectionist process. Indeed, outdated brands of electronics, candy, fashion, or music, which no longer compete in the marketplace, may nonetheless still influence how we feel about or understand the world. They affect the cultural evolution of worldviews despite no longer exerting any obvious impact on the external world. The human artifacts and behavior in existence at any time reveal some but not all aspects of the states of the culturally evolving worldviews that are generating them.

Some might fear that the proposed theory of culture necessitates accounting for an impossibly large number of variables in order to understand the mechanisms by which the collective information of a society evolves. Note that, in biology, individual traits are frequently considered as though they evolve independently from others, and this simplification gives considerable empirical power. The proposed theory does not stop one from doing experiments on isolated elements of culture such as words or gestures. Such experiments may tell us something about how culture evolves, though to the extent that they are based on simplifications they may give a distorted picture. When such elements of culture appear in the context of other such elements, their meanings change ways that are non-compositional; even simple conjunctions and disjunctions of words or concepts behave in ways that violate the rules of classical logic [97]. Moreover, to build a theory of cultural evolution based on such considerations is like looking for the quarter under the streetlamp knowing it is not there. The theory must attempt to capture the data completely and accurately even if the experiments we can currently perform do not do justice to it.

One avenue for future work in this direction involves application of reaction networks to cognitive and social phenomena [21]. The dynamics of the system—whether chemical, cognitive, or social—is described by a set of highly coupled non-linear differential equations. The qualitative behavior of the system is characterized in terms of phase states using notions from graph theory and algebraic analysis. Another avenue involves expanding formal models for describing the non-compositional manner in which concepts interact to generate complex conceptual aggregates such as ideas [44]. Two other avenues for future work are described in more detail in the following two sections.

### *6.1. EVOC: A computational model of cultural evolution*

One source of evidence for the above theory comes from a communal exchange based computational model of the EVolution Of Culture, abbreviated EVOC. Here it is described in only so much detail as is necessary; further details are provided elsewhere [29, 40, 41, 45, 46, 52, 78]. EVOC uses neural network-based agents that invent new ideas, imitate actions implemented by neighbors, evaluate ideas, and implement successful ideas as actions. In EVOC, the cultural representations are neither discrete nor continuous but distributed. Ideas for actions are distributed across mental representations of their body parts. Invention works by modifying a previously learned action, making use of hunches that build up over time about what makes for effective actions (such as that more overall movement, or symmetrical movement, tends to be good). The process of finding a neighbor to imitate works through a form of lazy (non-greedy) search. An imitating agent scans its neighbors, and adopts the first action that is fitter than the action it is currently implementing. If it does not find a neighbor that is executing a fitter action than the action it is currently implementing, it continues to execute that action. Over successive rounds of invention and imitation, agents' actions improve. EVOC thus models how cumulative adaptive change occurs in a cultural context. Agents do not evolve in a biological sense—they neither die nor have offspring—but do in a cultural sense, by generating and sharing ideas for actions.

We refer to the success of an action as its adaptive value, or simply, *value*. The adaptive landscape rewards head immobility and symmetrical limb movement.[10] The value of actions starts out low because initially all agents are immobile. Soon some agent invents an action that has a higher value than doing nothing, and this action gets imitated, so the mean adaptive value of actions across the artificial society increases over time as other ideas get invented, assessed, implemented as actions, and spread through imitation.

EVOC exhibits typical evolutionary patterns such as an initial increase in diversity as the space of possibilities is explored, followed by a decrease in diversity as agents converge on the best actions (Fig. 7). The mean value of actions for a society that uses imitation exceeds that of an a-cultural society that uses invention alone [29], a finding in agreement with results reported elsewhere [72].

[Insert Figure 7 here.]

The fact that EVOC generates cumulative, open-ended adaptive change indicates that a communal exchange theory of cultural evolution is computationally feasible. The theory is consistent with network-based approaches to modeling artifact lineages, trade, and the social exchange of knowledge and beliefs [2, 90]. It is also more consistent than a 'survival of the fittest' perspective on culture with findings that human society is more cooperative than is predicted by either expected utility or natural selection on genetic variation [9, 63] and with evidence that cultural evolution, construed in selectionist terms, actually reduces cooperation, at least in some situations[11] [77]. Thus there is theoretical, computational and empirical support for the communal exchange theory of cultural evolution.

A limitation of the current version of EVOC is that although the space of possible outputs is open-ended (there is no limit to the variety of possible outputs), it is open-ended in a highly constrained, almost trivial way, that reflects heuristics for generating novelty, rather than a conceptual network minimizing entropy by reconfiguring its connections. The EVOC code is being rewritten to be more consistent with the 'honing' view of creativity as the context-driven reconceptualization of an idea that may result when worldview self-organizes into a more stable, lower-energy state.

---

[10] This adaptive landscape was designed to meet several criteria. First, it exhibits a cultural analog of *epistasis*. In biological epistasis, the fitness conferred by the allele at one gene depends on which allele is present at another gene. In this cognitive context, epistasis is present when the fitness contributed by movement of one limb depends on what other limbs are doing. Thus the adaptive landscape enabled us to investigate how epistasis affects evolution in a cultural context. Second, because there are two sets of limbs (the arms and the legs) that could benefit from learning the trend that symmetrical movement tends to lead to fitter actions, it enabled us to investigate the effect of learning on evolution. Finally, since symmetrical limb movement is common in many actions it generates actions that are relatively realistic.

[11] For a review of the relationship between evolution and cooperation, see [139]. They argue that evolutionary concepts are widely misapplied in human research, and show that many of the reasons given for viewing human cooperation as special are contradicted by the biological facts. For example, humans are not unique with respect to (1) degree of altruism, (2) cooperation with non-relatives, or (3) the enforcement of cooperation through mechanisms such as punishment.

*6.2. Non-phylogenetic network-based approaches to documenting material cultural history*
EVOC attempts to capture the algorithmic structure of how culture evolves. It enables us to observe how various factors (such as ratio of inventors to imitators, population size or density, or media) impact the adaptive value and diversity of cultural elements. In contrast we now look at a computer model that encodes information about real human-made artifacts and takes a communal-exchange based approach to reconstructing historical lineages amongst them. Cultural phenomena that are analyzed using a phylogenetic approach are also amenable to an approach that is more consistent with the communal exchange theory of cultural evolution, such as Lipo's network-based approach [79]. This method, derived from graph theory, averts the problems discussed in Section 3.4 by ordering data according to similarity without necessarily implying common ancestry. Analysis of the data yields quite a different pattern of evolutionary change from that of selectionist-inspired approaches. Following O'Brien, samples that are rated the same with respect to all considered attributes are categories together as a particular taxon. Attributes are encoded as a number string. Each position in the string refers to a particular attribute, and the number at a position refers to the state of that attribute for the taxon (Fig. 8a).

[Insert Figure 7 here.]

Taxa are simply arranged according to the number of attributes by which they differ. The majority of taxa have two lines coming from them, one to a taxon that preceded it, and one to a taxon that followed it; the network does not specify which is which. Those that have more lines coming from them (such as the artifact labeled 31222122) reflect the existence of multiple other taxa with the same number of differences.

Several aspect of the procedure are noteworthy. First, the network-based approach does not make *a priori* assumptions about the sources of diversity. It is uncommitted with respect to whether differences reflect branching due to fission or blending due to transmission. Second, the method is also uncommitted with respect to chronology. Additional data indicate the directionality of the evolutionary pathway, as shown in Fig. 8b. Comparing this with the phylogenetic representation in Figure 2 reveals that a different pattern of ancestry is obtained.

The *conceptual network approach* is a computer model documenting material cultural history that was inspired by the communal exchange theory of cultural evolution. It builds on Lipo's [79] network-based approach, adding to it the capacity to incorporate reticulate relationships as well as hierarchical groupings, and incorporates conceptual information to complement physical attribute data. Thus, for example, it could incorporate that mortar and pestle are related by way of the concept 'complementarity' despite having few physical attributes in common.

The more superficial level of conceptual structure in the program consists of basic level concepts [104], such as PROJECTILE POINT, which incorporate tangible attributes of objects in the physical world. Although this basic level is the level at which items are first perceived, and at which we generally refer to and interact with them, the program allows one to work at a more subordinate conceptual level and thus consider a finer level of discrimination, *e.g*., FLUTED PROJECTILE POINT instead of PROJECTILE POINT. These concepts are pre-defined by the user, as opposed to being learned by the program by itself. Less superficial, more abstract levels of conceptual structure can also be defined, consisting of *superordinate concepts* such as WEAPON. Superordinate concepts often refer to multiple basic level categories (*e.g*., PROJECTILE POINT and KNIFE are both instances of WEAPON), and they are more general than the level at which we refer to and interact with items (*e.g*., different kinds of weapon are interacted with in different ways). The capacity to thereby incorporate hierarchical conceptual structure facilitates representation of objects and their interrelations as they are actually conceptualized.

Sometimes the structure of concepts derives from their history (how they were conceived in the past), and sometimes from other sources (*e.g*., horizontal transmission, adaptive modification, or copying error). Networks represent, not just taxa of artifacts, but relationships amongst them as they are conceived of in the minds of a particular population of individuals at a particular time and place. The tool collects meta-data for a set of known samples by asking the user questions about their presumed function and use. The questions are generated using a *conceptual network* that determines which questions are relevant for the sample. This leads to the creation of two networks: an attribute-level only one, and one that incorporates meta-data. The program analyses both the superficial attributes and abstract aspects of the samples, and uses this information to generate networks that show how the artifacts are most likely to have evolved chronologically. An early version of the conceptual network approach [47] was applied to the previously-mentioned set of early projectile points from the Southeastern United States that were modeled using a phylogenetic approach [93]. By incorporating not just superficial attributes of artifact samples (*e.g*., fluting) but also conceptual knowledge (*e.g*., information about function), a significantly different pattern of cultural ancestry emerged from that obtained using the phylogenetic approach (fig. 9).

[Insert Figure 9 here.]

Since using the entire data set generates output that is crowded and difficult to parse, the figure just shows a subset of the data. The output shows both the original Lipo Network (LN) approach and the Conceptual Network (CN) approach. In the LN approach, shown to the left, for any sample *x*, it is possible that more than one of the other samples is equally similar to *x, i.e*. minimizes the Hamming distance (the *N* function) with respect to *x*. Therefore, using attributes only, there is a large probability of generating the incorrect lineage. If you look to the samples featured on the upper right, it guessed that the terminal sample ‘Calfcreek’ is most closely related to the topmost sample, ‘Graham4’. Indeed based on the superficial attributes only, this was a reasonable guess.

A more recent version of the conceptual network model incorporates a means of varying the relative contributions of different types of data by incorporating a new parameter referred to as *perspective* [130]. The perspective parameter enables the user to define naturalistic groupings of attributes, and weight the extent to which these groups enter into the calculations of similarity used to generate lineages. For example, if one has more confidence in data concerning design features that are purely decorative or of symbolic importance then these attributes can be weighted more strongly than data on attributes that serve a functional purpose (Fig. 10).

[Insert Figure 10 here.]

When applied to the same Baltic psaltery data as used by Tëmkin and Eldredge [125], this version of the conceptual network model could incorporate not just physical characteristics but also hierarchically organized conceptual attributes pertaining to sacred symbolic imagery, and weight them differently. This resulted in the recovery of a previously unacknowledged pattern of historical relationship that is more congruent with geographical distribution and temporal data than is obtained with a cladistic approach (Fig. 11).

[Insert Figure 11 here.]

Thus when one takes into account not just *physical* traits, but traits with *conceptual* significance such as those that confer symbolic meaning, the inferred pattern of ancestry can be markedly different. These efforts, while preliminary, demonstrate that a communal exchange based approach to documenting cultural history is not only possible, but avoids problems encountered with approaches inspired by a selectionist view of cultural evolution. Again, the point here is not that competition is not a significant component of cultural change; clearly groups with superior weapons or organization have eliminated other groups, and competition amongst groups has played a significant role in human history. The point here is that inter-group cultural competition is not the competition amongst instruction sets for self-replicating structures that characterizes a selectionist process.

*6.3. Meeting the challenges for a theory of cultural evolution*

Let us now return to the challenges put forward for a theory of cultural evolution at the beginning of the article. A first challenge was to explain why humans alone have evolved complex, cumulative, open-ended culture. It has been proposed that what is unique to the human species, and what enabled human culture to become an *evolutionary* process, was onset of not the ability to imitate or influence others but the capacity to take an idea and put one's own spin on it, reframing it in one's own terms [33, 39]. Communal exchange theory provides a natural explanation for how this could come about. Recall that in Kauffman's computer simulations of the origin of life through autocatalytic closure, each polymer was composed of up to a maximum of $M$ monomers, and assigned a low a priori random probability $P$ of catalyzing each reaction. The lower the value of $P$, the greater $M$ had to be, and vice versa in order for autocatalytic closure to occur, and $P$ could not be too high. A similar trade-off is expected between the cognitive analog of $M$, the maximum number of dimensions along which episodes are encoded, and the cognitive analog of $P$, the probability for items in memory to evoke one another [32]. So long as $\{M, P\}$ are sufficiently high, by engaging in self-sustained streams of thought that revise one's web of understandings, conceptual closure is re-established. Once there is a sufficiently high density of associations such that each experience can be re-framed in one's own terms, ideas can be adapted to the individual's particular needs, tastes, and desires, and thereby evolve. Thus communal exchange theory leads to the hypothesis that what enabled complex, cumulative, open-ended culture to evolve is onset of the capacity for *recursive* *recall,* in which one thought evokes another and so forth in a chain of associations. This enabled humans to respond to dissonance, frustration, or misunderstanding by considering from different perspectives and adapting ideas to new contexts, thereby reducing entropy and re-establishing conceptual closure. In other words, the 'hub' of cultural evolution should be a cognitive structure that generates transmittable novelty through this kind of self-mending process. Communal exchange theory further predicts that that newly discovered or artificially created life forms should be able to evolve culture if they possess cognitive structure that (as a result of the capacity for self-mending) generates transmittable novelty. A selectionist theory of culture is incompatible with the hypothesis that recursive recall plays a pivotal role in the onset of a cultural evolutionary process because to the extent that elements of culture are modified between transmission events (*e.g*., the adapting and re-thinking of ideas within the minds of individuals, through processes such as recursive recall, before they are transmitted to others), change cannot be due to differential selection of variants but to the factors governing these between-transmission processes.

This analysis suggests that other species are not prohibited from evolving complex cognition *a priori*, but that without the physical capacity to generate and manipulate complex artifacts and vocalizations there is insufficient evolutionary pressure to increase brain size sufficiently to obtain a sufficiently high $M$, and to tinker with $P$, until they achieve the requisite delicate balance to sustain a closure restoring stream of thought. Humans may be the only species (we know of) for which the benefits of this tinkering process have outweighed the risks. This is all the more reasonable because with the capacity for complex speech and hand movement there are clearly more ways to manifest complex thoughts*, i.e*., more ways in which they could yield outcomes in the world. To know precisely why conceptual closure arose in humans alone, further research is needed into how physical characteristics constrain and enable the evolution of cognitive abilities.

## 7. Conclusions and future directions

Given the current accelerated pace of cultural change, and its transformative effects on ourselves and our planet, it is becoming imperative to obtain a solid understanding of how culture evolves. Since cultural evolution, like biological evolution, results in the generation of cumulative, open-ended, adaptive novelty, it would appear to make sense to draw upon biological evolutionary theory to explain cultural evolution.

The propagation of Dawkins' view of natural selection, a view that does not capture the basic algorithmic structure of the process has led to efforts to frame cultural evolution in Darwinian terms. Examining the question of whether elements of culture constitute replicators, we saw that Dawkins' notion of replicators captures certain aspects of self-replication but ignores some of the most fundamental principles of natural selection. These include: (1) a code that gets used both as self-assembly instructions and self-description, (2) a genotype/phenotype distinction, and (3) a lack of transmission of acquired characteristics, where (2) and (3) are a consequence of (1). The assumption that the replicator notion captures the essential elements of natural selection has misleadingly led to support for the idea that culture evolves through a selectionist process.

Natural selection explains change in the frequency of inherited traits (traits transmitted from one generation to another by way of a self-assembly code such as the genetic code), not acquired traits (traits obtained between transmission events). A selectionist model is applicable when there is negligible transmission of acquired characteristics. If a process involves transmission of acquired traits and/or non-random variation, evolution over time is explained by whatever is biasing or modifying the variation, not by natural selection, *i.e.*, it is not due to statistical change in the frequency of heritable variations over generations due to differential response to selective pressure. Thus attempts to force culture into a Darwinian framework (even those that pay lip service to the importance of a "rich psychology") leads to models wherein "psychology" is reduced to sources of variation that occur during social transmission through processes such as copying error. Failure to appreciate Darwinian approaches to culture do not stem from a "tendency to think categorically rather than quantitatively" (p. 134) as Henrich et al. [60] claim, nor are critics "not well equipped to digest mathematical models" (p. 121). Criticisms of these models reflect genuine limitations in the methods used, and the theoretical foundation upon which they are based. Nevertheless, Darwinian approaches to culture such as memetics and dual inheritance theory have made an important contribution toward a viable scientific framework for culture by aiding identification of precisely where the analogy between biology and culture breaks down.

More recent efforts to frame culture in evolutionary terms, which take their cue from communal-exchanged based theories of the origin of life, suggest that non-Darwinian, epigenetic aspects of biological evolution that may be useful in developing an evolutionary framework for culture. This direction is in line with emerging efforts (such as can be found in [17]) to bridge the gap between evolutionary approaches to anthropology and archaeology on the one hand, and interpretive and social constructionist approaches on the other. Research toward a communal exchange theory of cultural evolution is focused on computational models that incorporate empirical findings about how humans creatively adapt ideas to new situations.

In sum, recent decades have witnessed considerable progress toward determining in what sense culture evolves. There is reason to believe that we are on our way to a scientific framework for the process by which, by building cumulatively on each other's ideas and applying them in new contexts, humans have transformed our planet.

**Acknowledgements**
This research was conducted with the assistance of grants from the National Science and Engineering Research Council of Canada, and the Fund for Scientific Research of Flanders, Belgium.

**Table 1.**

| | **Replicator** | **Self-replicating Automaton** |
|---|---|---|
| Self-replication | Yes | Yes |
| Longevity; fecundity; fidelity | Yes | Yes |
| *Passive copying and active transcription of self-assembly instructions* | ? | Yes |
| *Sequestration of inherited information* | ? | Yes |
| Genotype / phenotype distinction | ? | Yes |
| Transmission of acquired traits | ? | No |
| Evolutionary processes it seeks to explain | Biological; cultural | Biological |

Table 1. Comparison of replicators and self-replicating automata. Both involve self-replication with longevity, fecundity, and fidelity. Only the self-replicating automaton requires *self-assembly instructions that are both passively copied and actively transcribed.* Therefore, only the self-replicating automaton is committed to (1) *sequestration of inherited information, (2) clear* distinction between genotype and phenotype, and (3) prohibition on transmission of acquired traits. The replicator has been proposed as the central construct of an evolutionary framework for both biological evolution (of genes) and cultural evolution (of memes). Since cultural evolution lacks *self-assembly instructions that are passively copied and actively transcribed, and* transmission of acquired traits is not prohibited, the self-replicating automaton has been proposed as the central construct for biological evolution only.

**Table 2.**

| | **Natural Selection** | **Communal Exchange** |
|---|---|---|
| Unit of self-replication | DNA | Autopoietic network |
| Mechanism for preserving continuity | DNA replication; proofreading enzymes, etc. | Retention of horizontally transmitted information |
| Mechanism for generating novelty | Mutation; recombination; replication errors; pseudo-genes | Faulty duplication of autopoietic structure; transmission errors; innovation |
| Self-assembly code | Yes | No |
| High fidelity | Yes | No |
| Transmission of acquired traits | No | Yes |
| Type of process | Darwinian | Lamarckian (by some standards) |
| Evolutionary processes it seeks to explain | Biological | Early life; horizontal gene transfer (HGT); cultural |

Table 2. Similarities and differences between two evolutionary frameworks: natural selection and communal exchange. Both have mechanisms for preserving continuity and introducing novelty. However, whereas natural selection is a high fidelity Darwinian process and the self-replicating structure is DNA-based self-assembly instructions, communal exchange is a low fidelity Lamarckian process, and the self-replicating structure is an autopoietic network. Only communal exchange allows transmission of acquired traits. Communal exchange is proposed to be the mechanism underlying the evolution of early life, culture, and some aspects of present day life, *e.g.*, horizontal gene transfer.

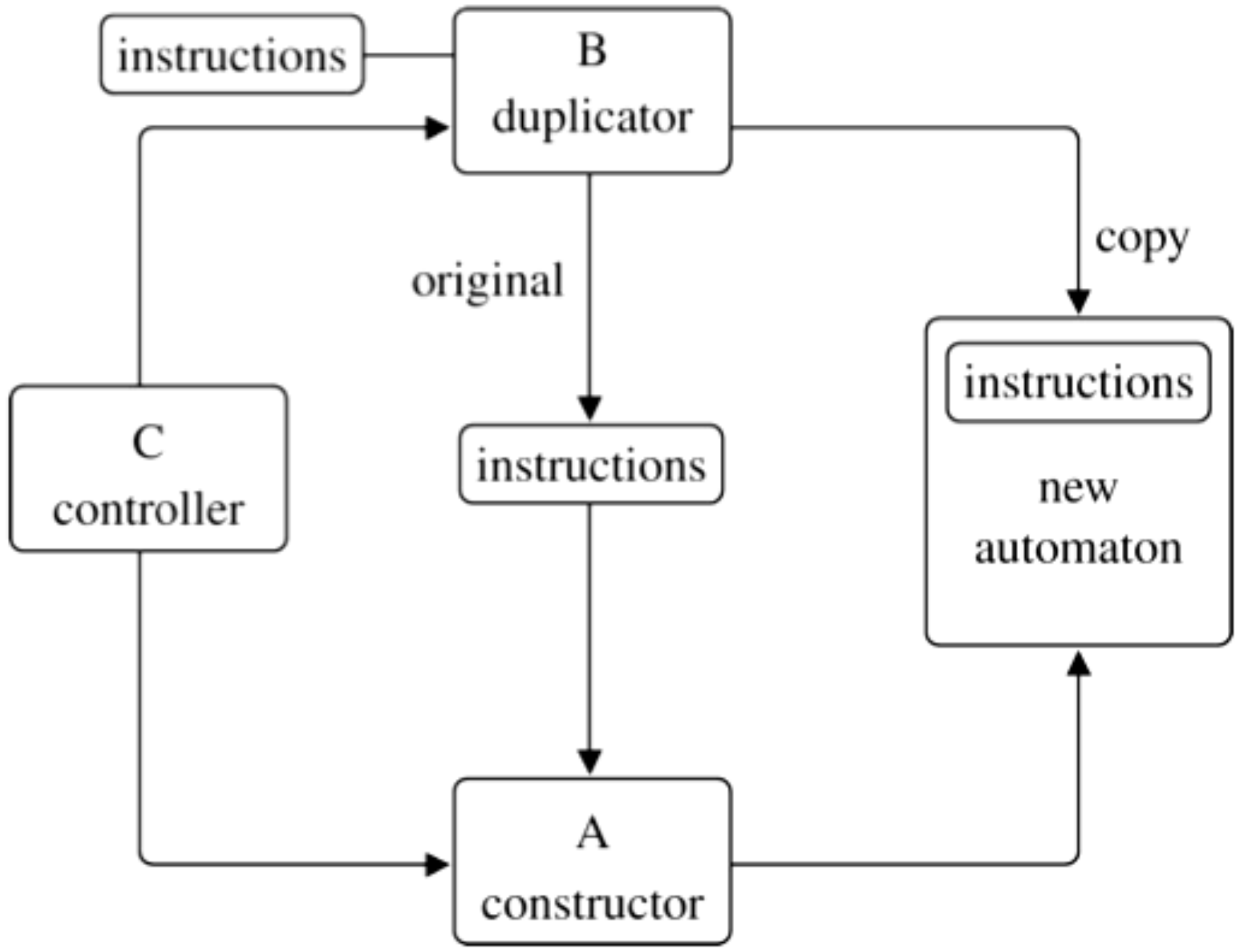

Figure 1. A self-replicating automaton. von Neumann determined that a physical structure capable of self-replication consists of a constructor (A), which can build a new system using material in the surroundings, an instruction set (such as DNA) containing information about what must be built by the constructor, a duplicator (B) which copies the instructions, and a controller (C) which ensures that these events takes place in a well-defined sequence. He referred to this as a *self-replicating automaton*. (Adapted from [132].)

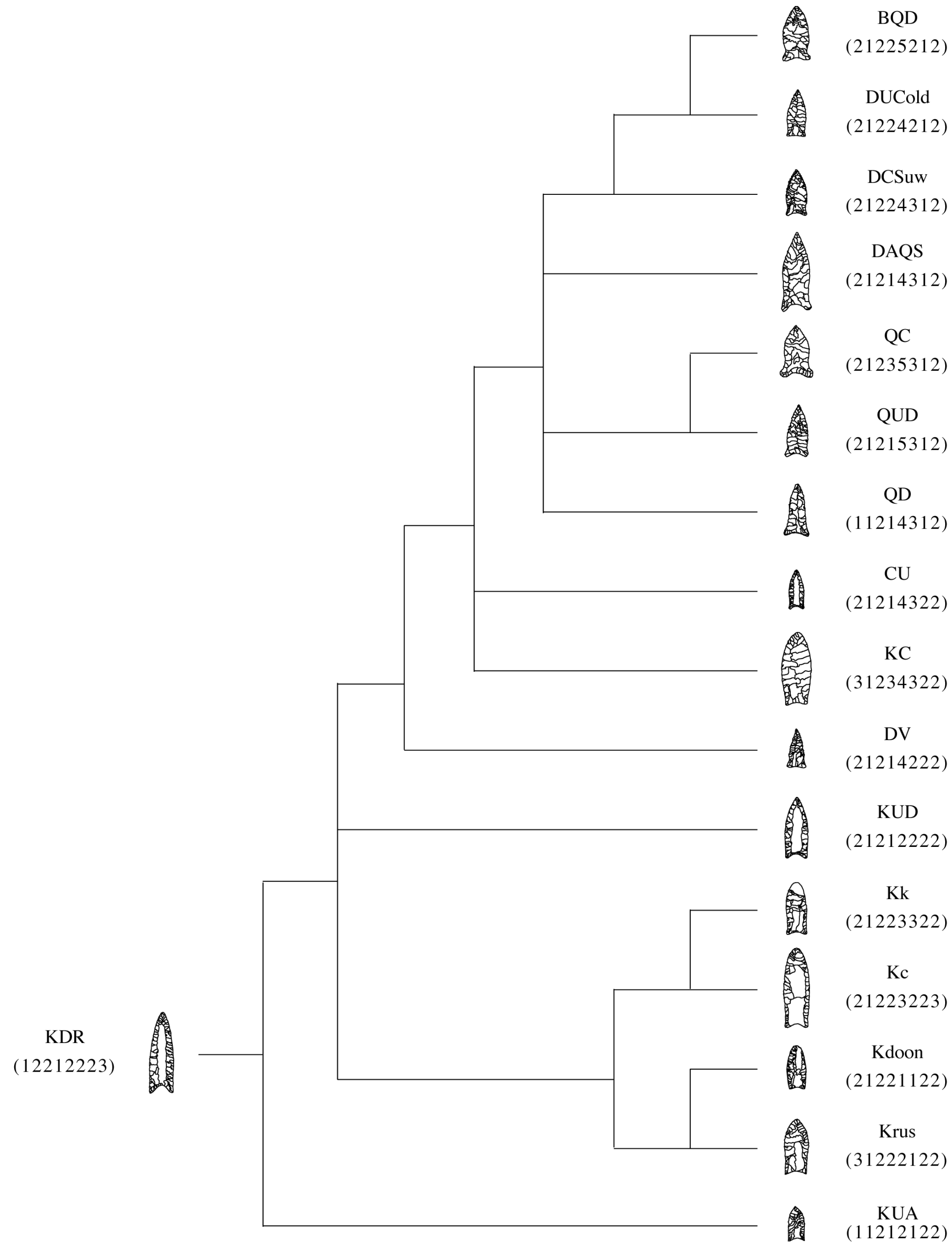

Figure 2. Phylogenetic representation of PaleoIndian period projectile points from the Southeastern United States with 17 taxa defined by 18 attributes. From [93].

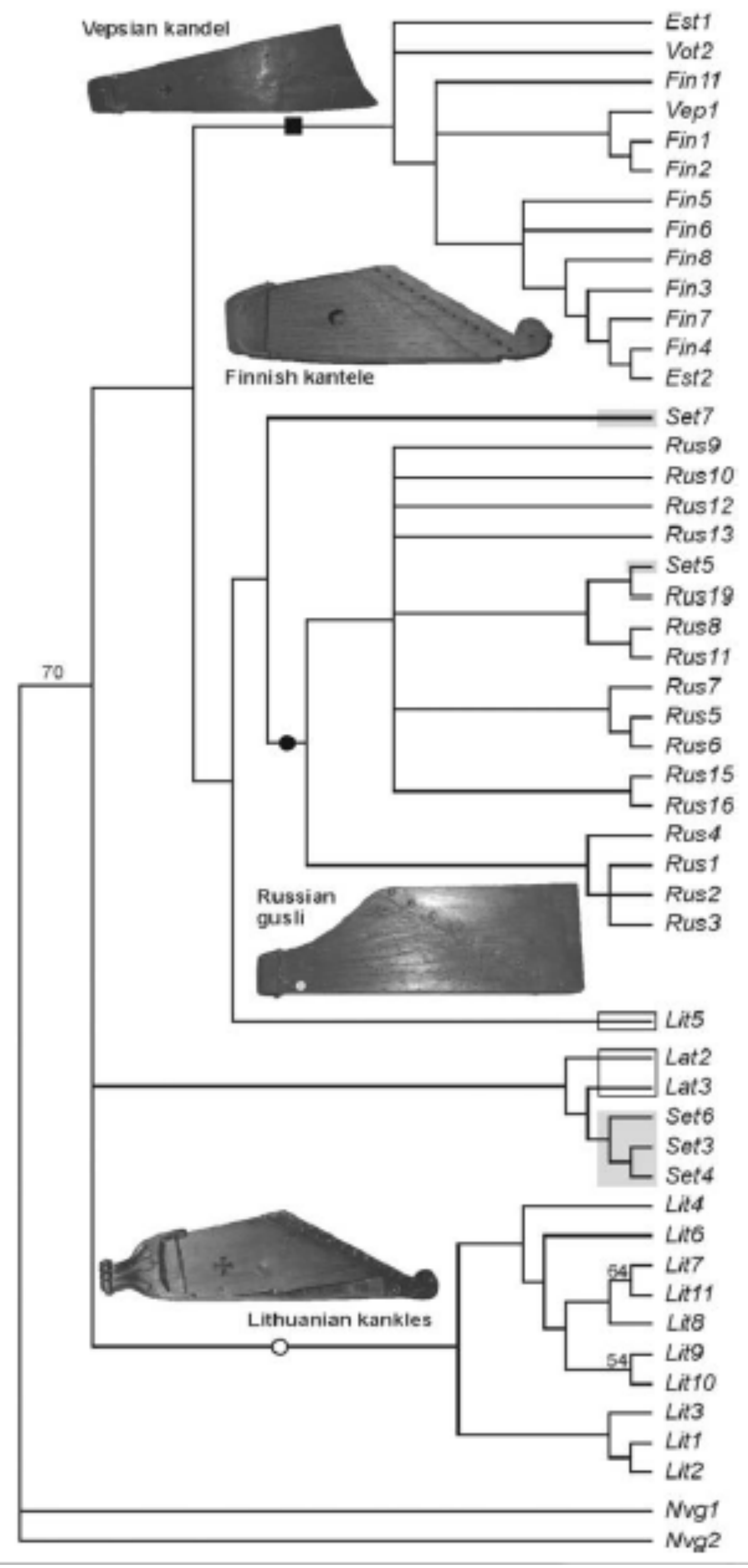

Figure 3. Phylogenetic representation of the evolution of the Baltic Psaltery. Note that similarity is assumed to derive from common ancestor (thus the branching pattern) as opposed to horizontal exchange (*e.g.*, through imitation or trade). From [125].

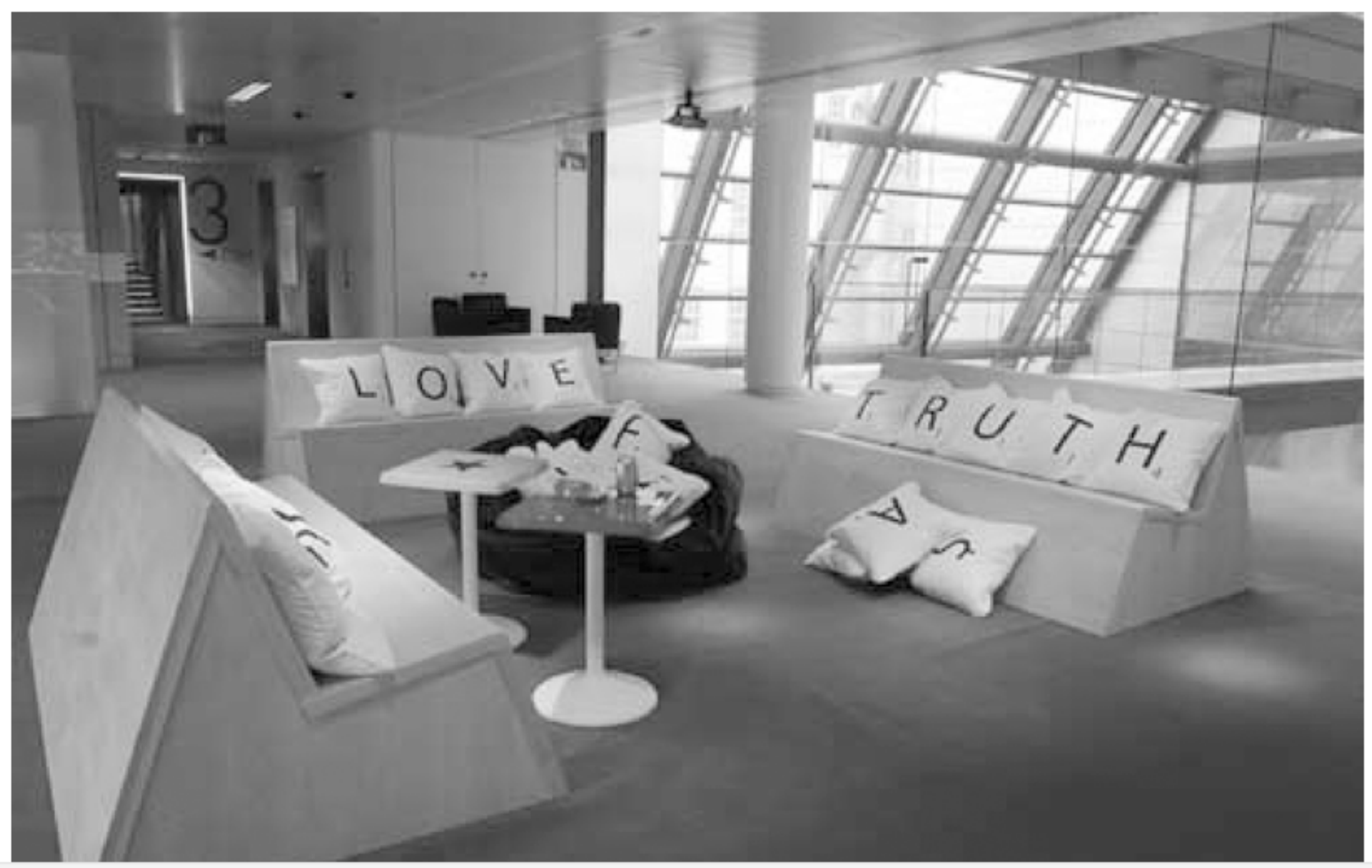

Figure 4. This furniture inspired by the game of scrabble is an example of the ubiquitous phenomenon of blending of cultural 'lineages'.

Evolution Through Natural Selection

**Lineage A**

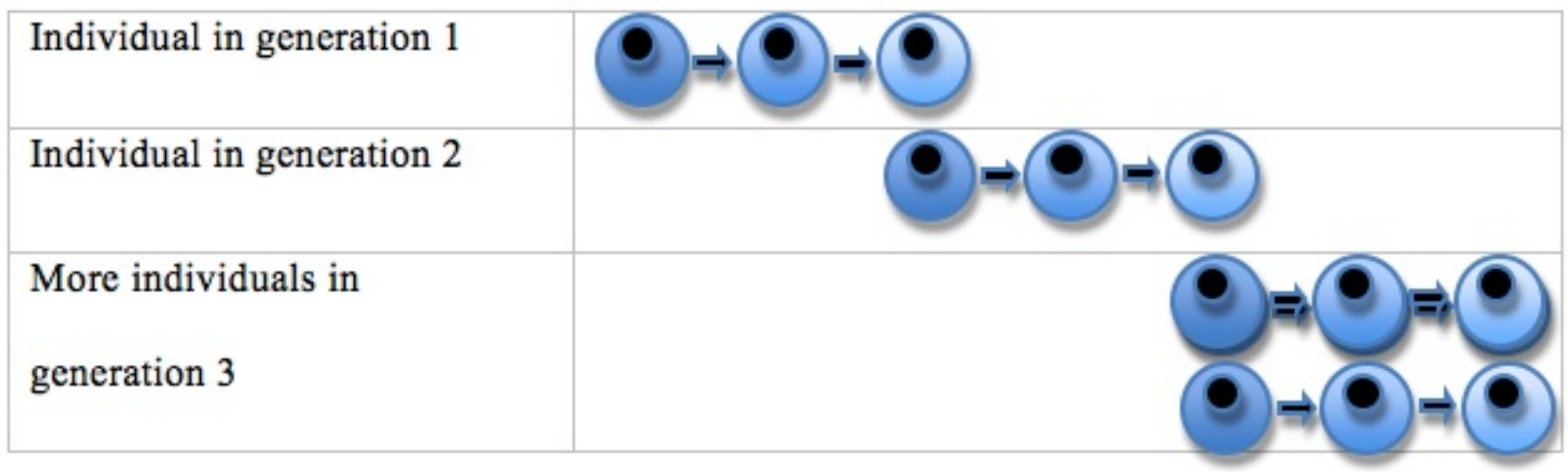

| Individual in generation 1 | |
|---|---|
| Individual in generation 2 | |
| More individuals in generation 3 | |

**Lineage B**

| Individual in generation 1 | |
|---|---|
| Individual in generation 2 | |
| No individuals in generation 3 | |

Evolution Through Communal Exchange

**Lineage A**

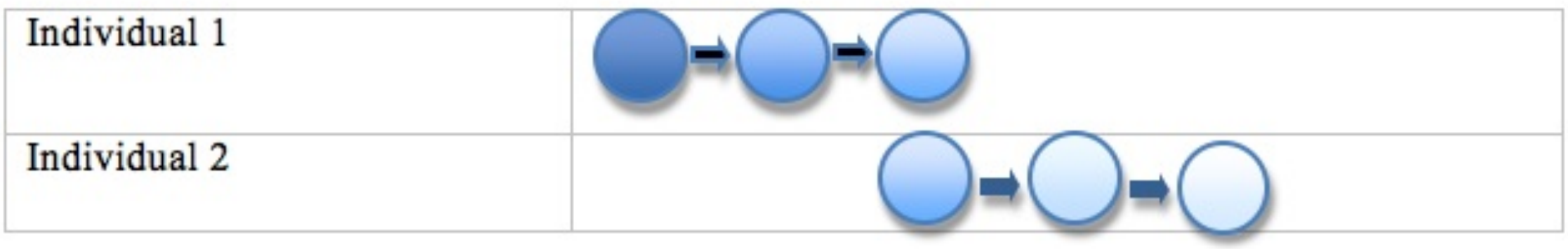

| Individual 1 | |
|---|---|
| Individual 2 | |

**Lineage B**

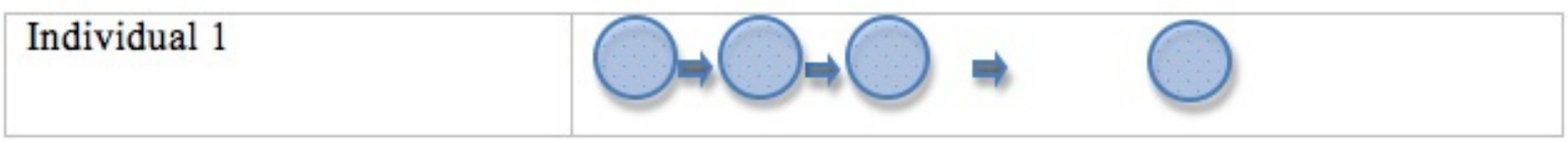

| Individual 1 | |
|---|---|

Figure 5. Schematic depiction of evolution through natural selection, top, occurs through competition amongst genetic variants. Each individual has a genotype, represented by a small circle, and a phenotype, represented by a larger circle encompassing the small circle. Phenotypic change acquired over a lifetime, indicated by increasingly lighter color, is not passed on, as indicated by the fact that in lineage A the individual born in generation 2 did not inherit the lighter color the parent acquired by the time it had offspring. Lineage A outcompetes lineage B as indicated by the fact that by generation 3 there are no individuals left in lineage B. Evolution relies more on competition amongst individuals than transformation of individuals.

A more primitive process, evolution through communal exchange, bottom, occurs through duplication and transformation of variants. There is no distinction between phenotype and genotype, and acquired characteristics are transmitted, as shown by the fact that in lineage A, Individual 2 has the light color its 'parent' acquired over its lifetime. Death does not have the same finality, since given the right conditions an inert individual could potentially reconstitute itself, a situation represented by Lineage B. Therefore the term 'generations' is not meaningful. Changes acquired over a lifetime are transmitted. Evolution relies more on transformation of individuals than on competition amongst individuals. Biological evolution involves both communal exchange and natural selection. It is posited here that culture evolves through communal exchange.

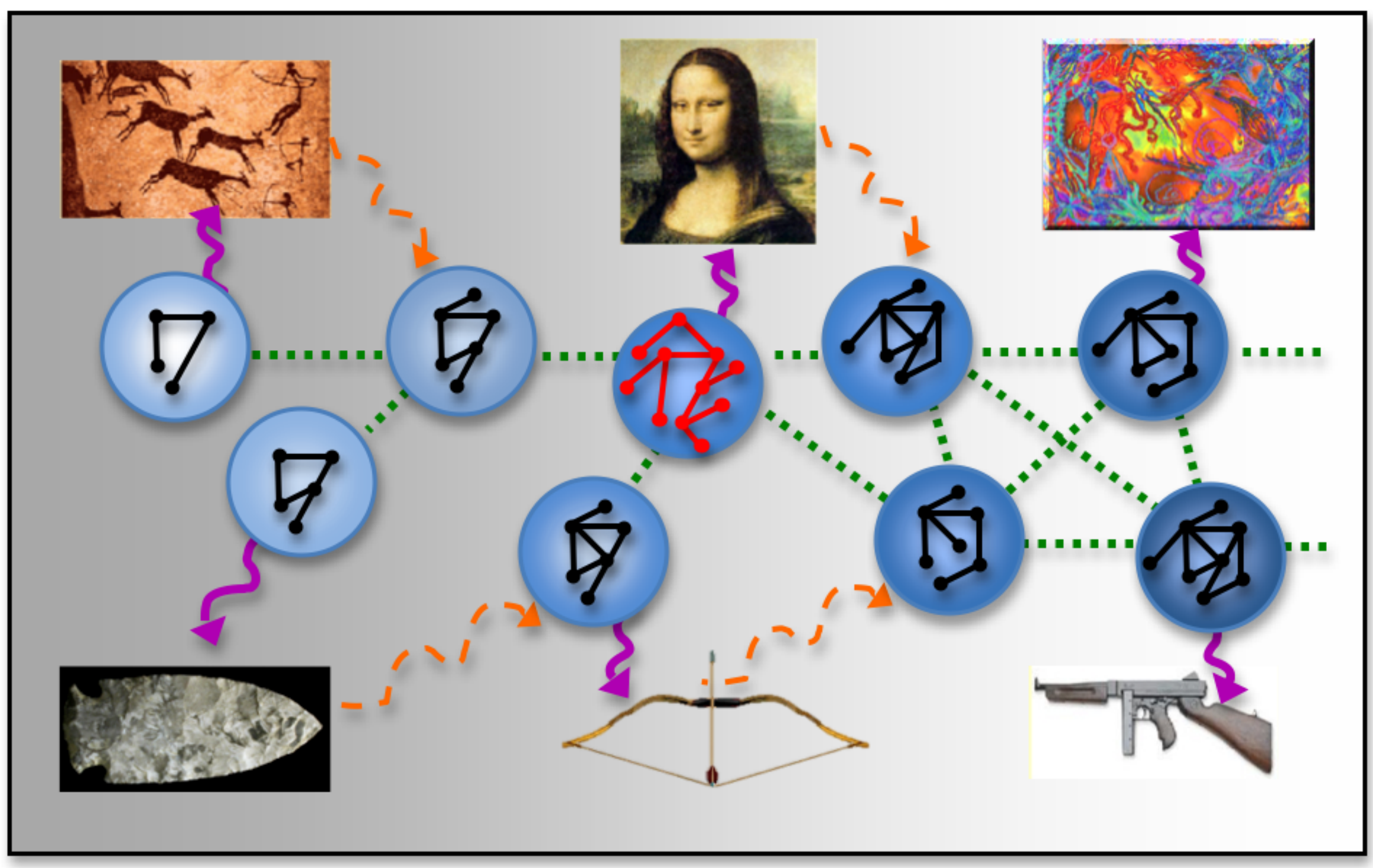

Figure 6. Schematic depiction of how worldviews transform through communal exchange, and thereby evolve not through *survival of the fittest* but *transformation of all.* Individuals are represented by spheres, and their internal models of the world, or *worldviews*, are represented by networks within the spheres. Patterns of social transmission are indicated by dashed green lines. Creative contributions to culture are indicated by wavy purple arrows from creator to artifact. Learning through exposure to artifacts is indicated by wavy orange arrows. Worldviews and patterns of social transmission tend to become more complex over time. Individuals such as the one with the red network are more compelled than others to reframe what they learn in their own terms, potentially resulting in a more unique and/or nuanced worldview. Such 'self-made' individuals are more likely to exert transformative effects on the world such as through creation of artifacts, which may influence the formation of new worldviews long after they have died, as indicated by the proliferation of segments of red network in individuals exposed to the Mona Lisa or its creator.

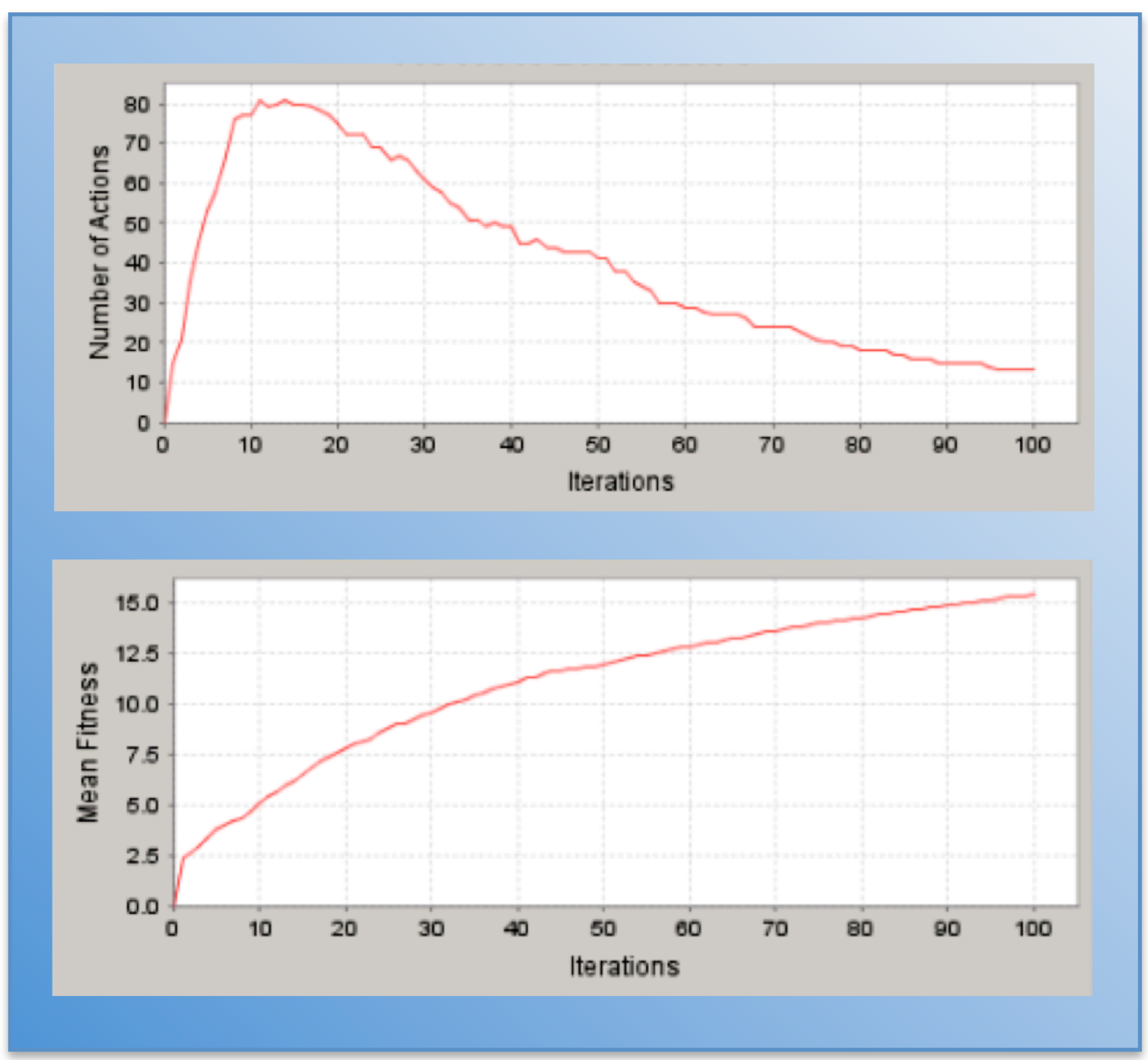

Figure 7. Output of a computational model of cultural evolution composed of 100 neural network based agents that invent ideas for actions by modifying old ones and communally exchanging ideas by imitating neighbors. Note that for each adaptive landscape there are at least hundreds (and in some cases an infinite number) of possible cultural variants, which makes it possible to model not just cultural transmission but the adaptive evolution process that gives rise to increasingly fit variants over time. This stands in sharp contrast with dual inheritance models and other selectionist models, which consider as few as two cultural variants. The adaptive value of actions increases throughout the run (top). Diversity increases as the space of possibilities is initially explored, and decreases as agents converge on the fittest actions (bottom). The incorporation of not just superficial attributes of artifact samples (*e.g.*, fluting) but also conceptual knowledge (*e.g.*, information about function), gives a significantly different pattern of cultural ancestry emerged from that obtained using the phylogenetic approach.

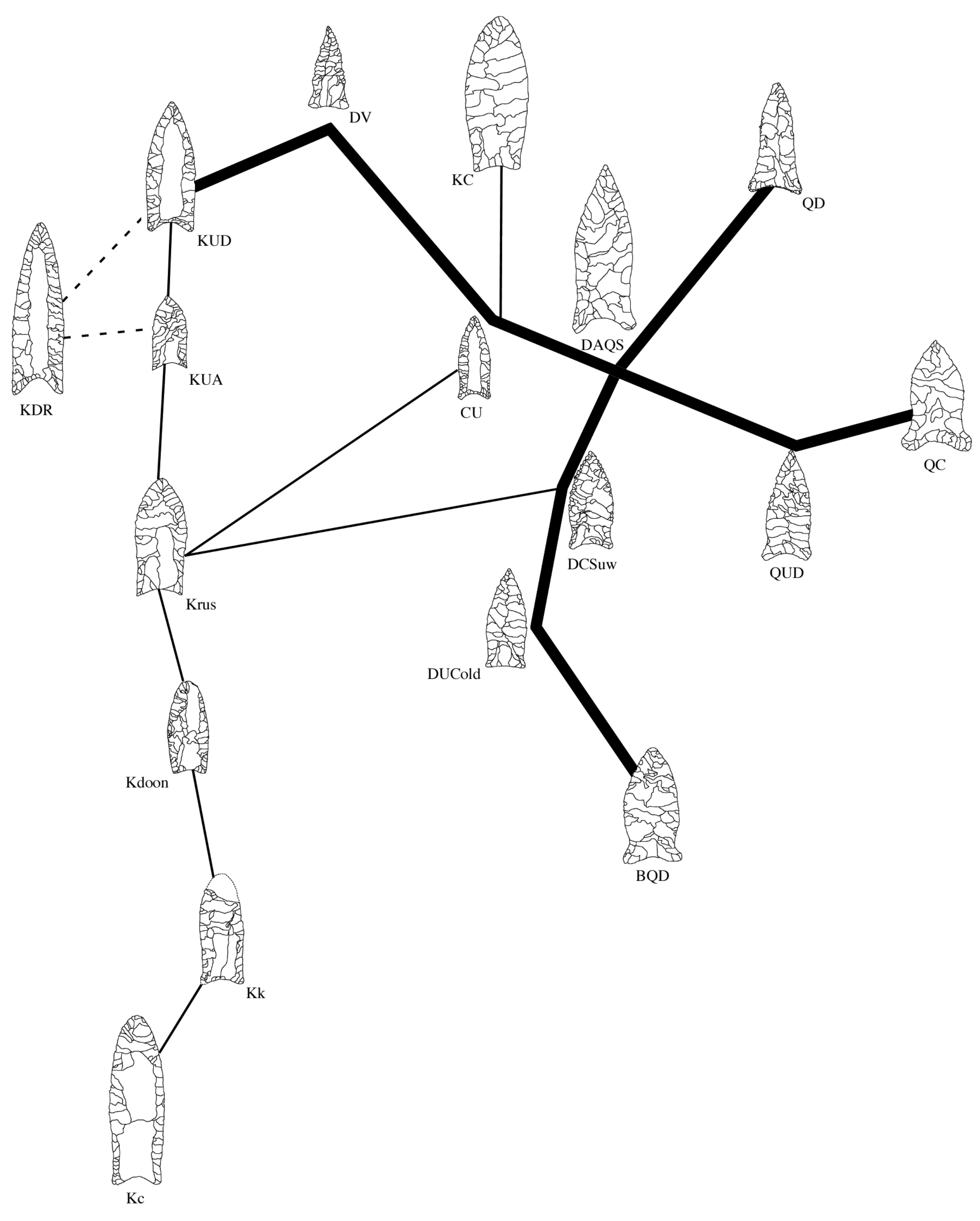

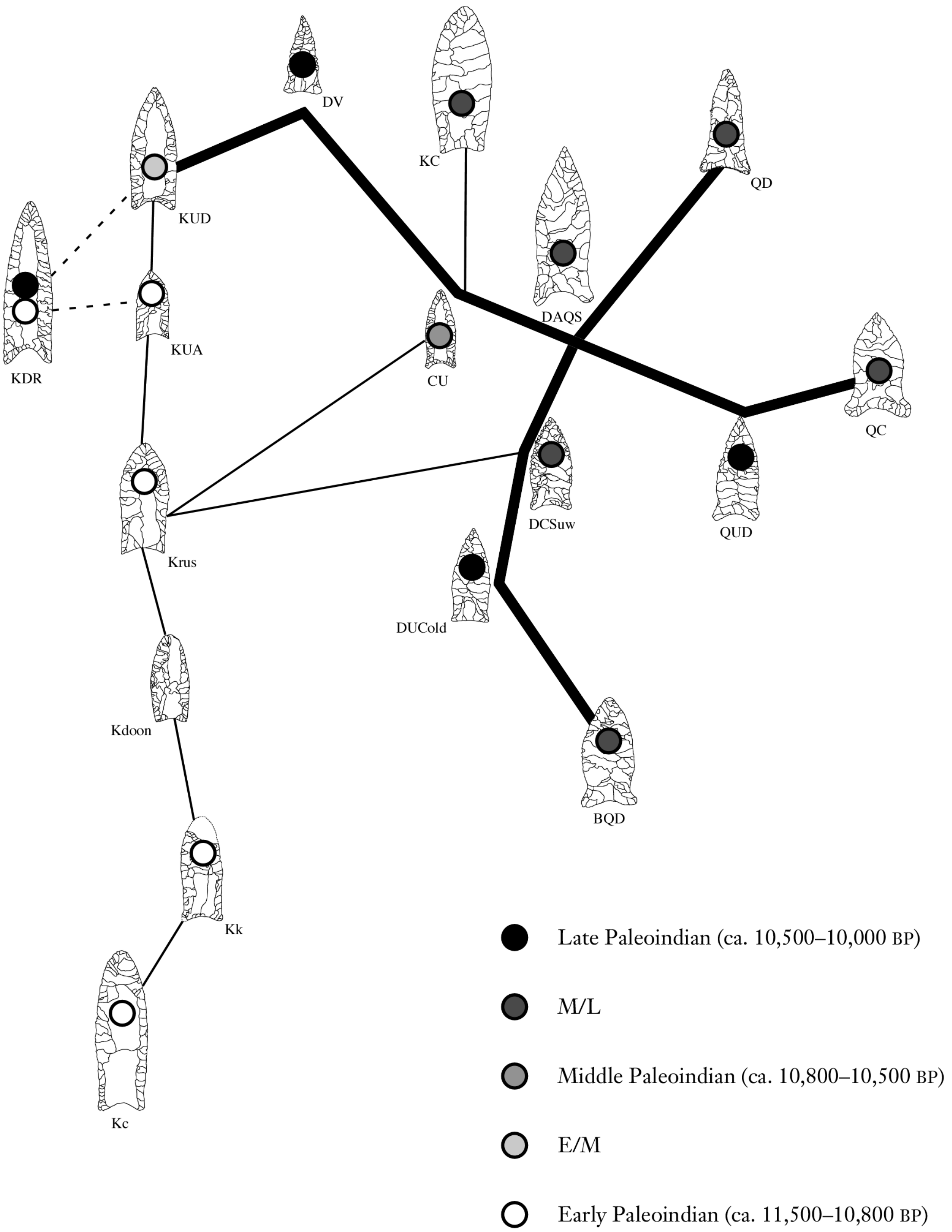

Figure 8. (a) Top: Graph produced by linking taxa to their most similar neighbors. Bold lines represent differences of only one attribute. Thin solid lines show differences of two attributes. Dotted lines show differences of three attributes. The multiple lines connecting taxon 31222122 to other taxa indicate ambiguity due to equivalent number of differences between multiple taxa. (b) Bottom: Graphical analysis of projectile point data with temporal information indicated by degree of shading of circles. From [79].

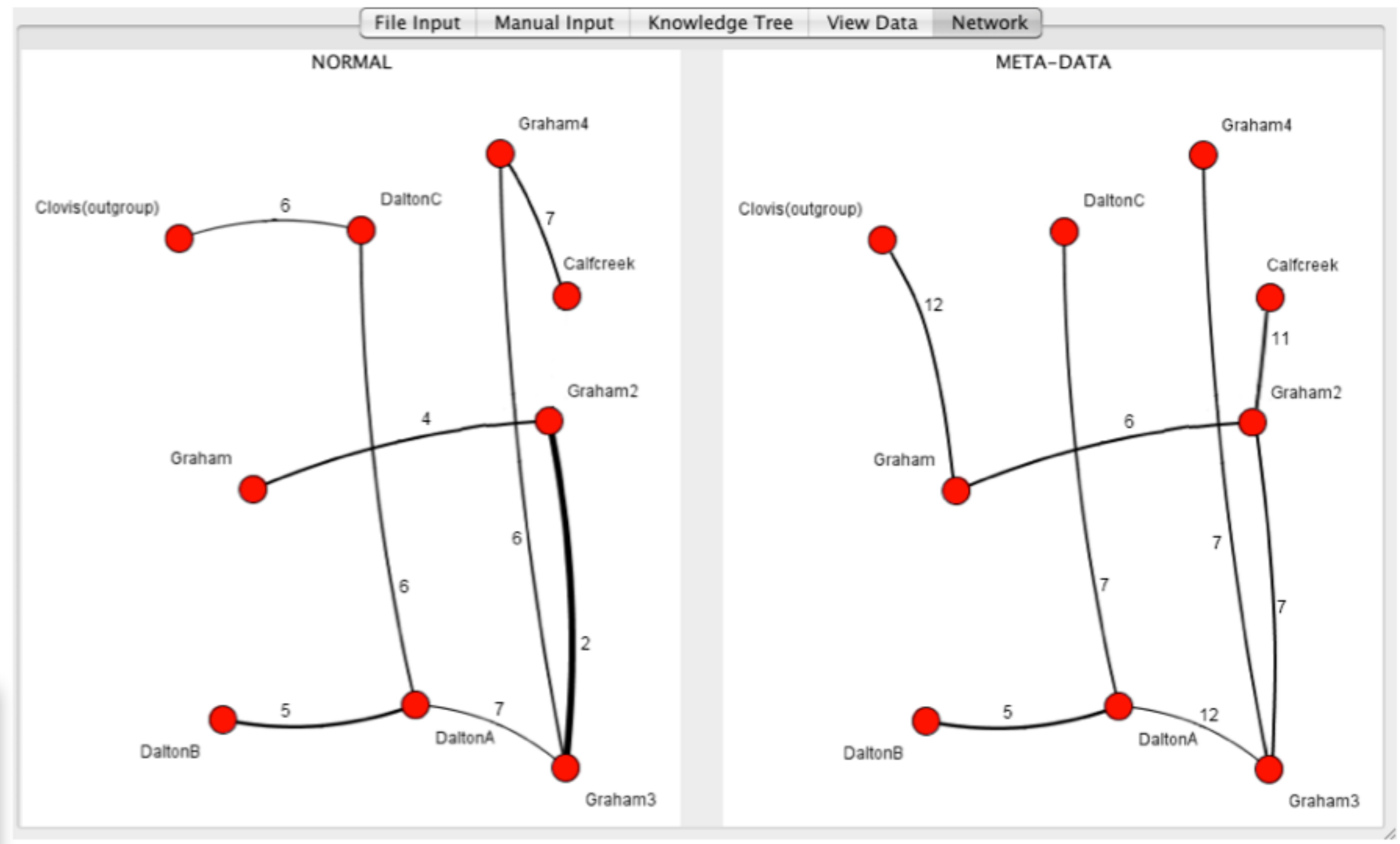

Figure 9. Two examples of conceptual network output given the same input data. Circles represent particular samples. Numbered lines give estimates of relatedness (lower numbers more closely related). The output on the left makes use of superficial attributes only. The output on the right additionally makes use of conceptual meta-data. From [47].

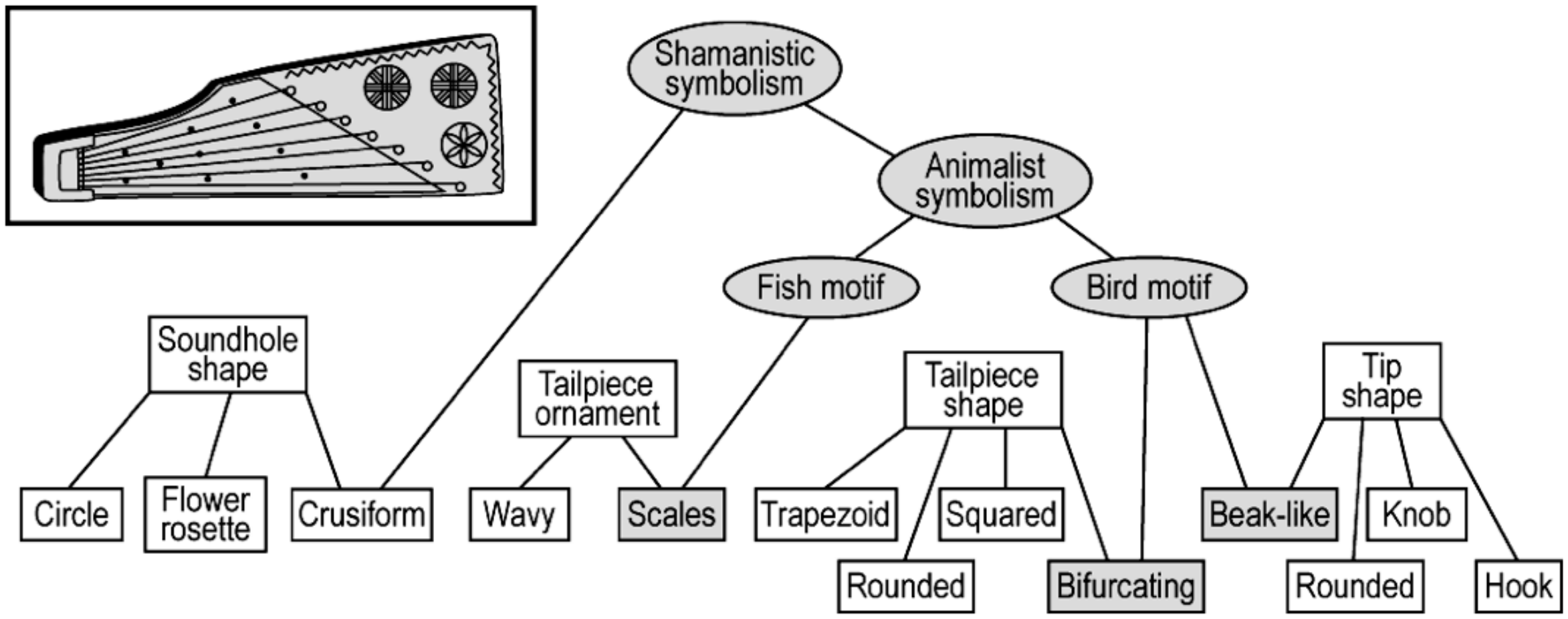

Figure 10. A segment of conceptual structure used in Baltic psaltery analysis. Elements in square boxes represent physical attributes of artifacts. Elements in oval boxes correspond to conceptual attributes. Shaded boxes designate both physical and conceptual attributes associated with symbolic significance, *i.e.*, that define the "Symbolism" perspective. Inset shows structure of a Baltic psaltery. From [130].

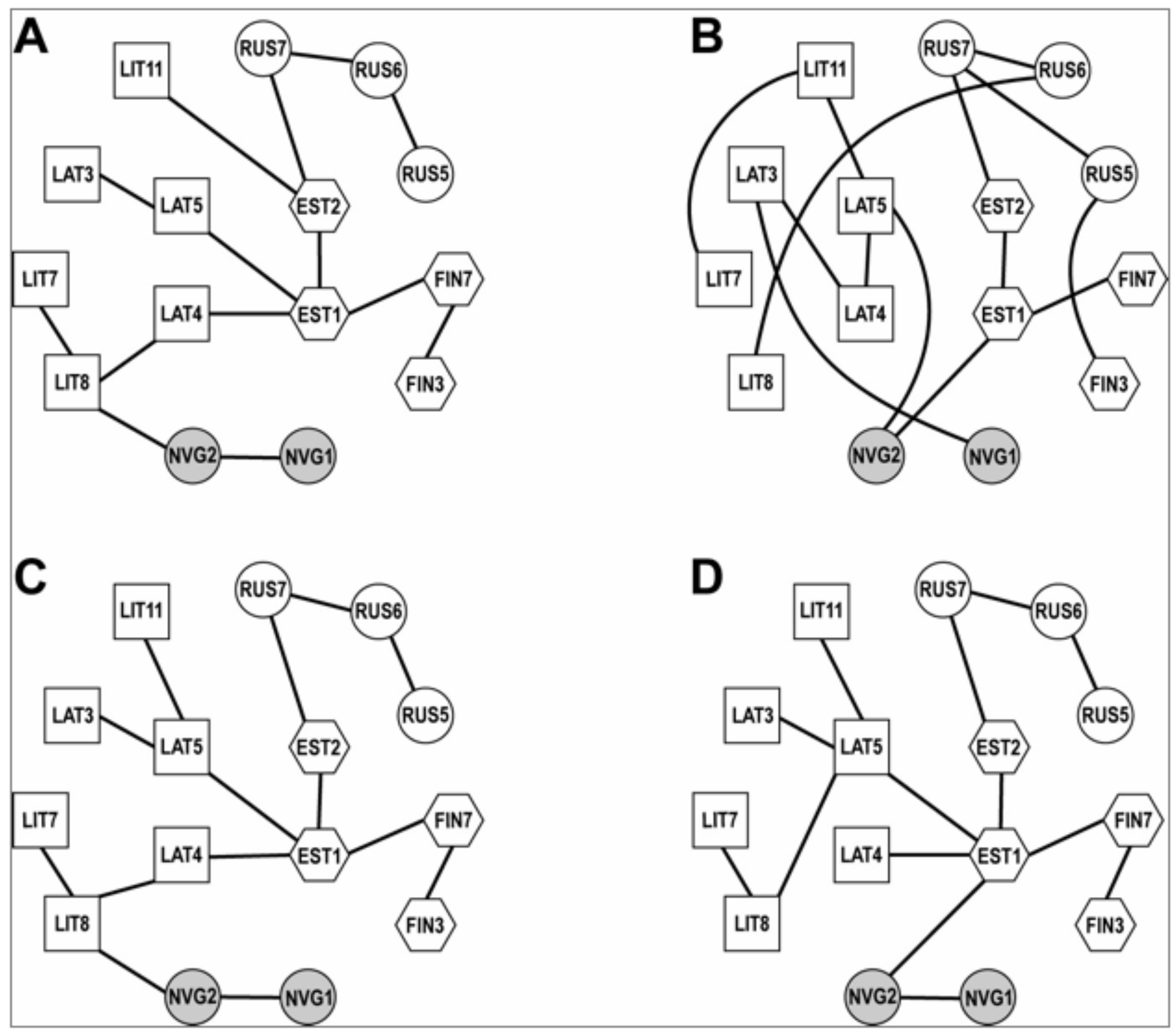

Figure 11. Similarity graphs based on conceptual network analysis of Baltic psalteries under different perspective weighting schemes. (A) Physical attributes only; (B) Symbolism; (C) Physical attributes and Symbolism (equal weights); (D) Physical attributes (25% weight) and Symbolism (75% weight). Each node corresponds to a single artifact. Node shapes indicate ethnolinguistic groups: Slavic (circle), Finnic (hexagon), and Baltic (square). Shaded nodes designate archaeological instruments (10-13 cc); remaining nodes correspond to ethnographical instruments (17-20 cc). From [130]. By incorporating not just physical characteristics but also conceptual attributes (such as those pertaining to sacred symbolic imagery, and weight them differently, it was possible to resolve ambiguities obtained with the phylogenetic approach, and generated a lineage more consistent with other historical data.